\documentclass[journal]{IEEEtran}
\usepackage{cite}
\usepackage{graphicx}
\usepackage{amsmath,amssymb,amsfonts}
\usepackage{algorithm, algorithmic}
\usepackage{subfigure}
\usepackage{multirow}
\usepackage{array}
\usepackage{color}
\usepackage{verbatim}
\usepackage{CJKutf8}
\usepackage[normalem]{ulem}

\begin{document}

\title{Deep Joint Source-Channel Coding for Image Transmission with Visual Protection}
%
%
%

\author{Jialong Xu,~\IEEEmembership{Member,~IEEE,}
        Bo Ai,~\IEEEmembership{Fellow,~IEEE,}
        Wei Chen,~\IEEEmembership{Senior Member,~IEEE,}
        Ning Wang,~\IEEEmembership{Member,~IEEE,}
        and~Miguel Rodrigues,~\IEEEmembership{Fellow,~IEEE}
\thanks{This work is supported by the Natural Science Foundation of China (62122012, 62221001); the Beijing Natural Science Foundation (L202019, L211012); the Fundamental Research Funds for the Central Universities (2022JBQY004). \textit{(corresponding authors: Bo Ai; Wei Chen)}}
\thanks{Jialong Xu is with State Key Laboratory of Advanced Rail Autonomous Operation, Beijing Jiaotong University, China, and also with Frontiers Science Center for Smart High-speed Railway System, China (e-mail: jialongxu@bjtu.edu.cn).}
\thanks{Bo Ai is with State Key Laboratory of Advanced Rail Autonomous Operation, Beijing Jiaotong University, China, also with Beijing Engineering Research Center of High-speed Railway Broadband Mobile Communications, China, and also with School of Information Engineering, Zhengzhou University, China (e-mail: boai@bjtu.edu.cn).}
\thanks{Wei Chen is with State Key Laboratory of Advanced Rail Autonomous Operation, Beijing Jiaotong University, China, and also with Key Laboratory of Railway Industry of Broadband Mobile Information Communications, China (e-mail: weich@bjtu.edu.cn).}
\thanks{Ning Wang is with School of Information Engineering, Zhengzhou University, China (e-mail: ienwang@zzu.edu.cn).}
\thanks{Miguel Rodrigues is with the Department of Electronic and Electrical Engineering, University College London, London, WC1E 7JE, U.K. (e-mail: m.rodrigues@ucl.ac.uk).}
}

\markboth{Journal of \LaTeX\ Class Files,~Vol.~14, No.~8, August~2022}%
{Xu \MakeLowercase{\textit{et al.}}: Deep Joint Source-Channel Coding for Image Transmission with Visual Protection}

\maketitle

\begin{abstract}
Joint source-channel coding (JSCC) has achieved great success due to the introduction of deep learning (DL). Compared to traditional separate source-channel coding (SSCC) schemes, the advantages of DL-based JSCC (DJSCC) include high spectrum efficiency, high reconstruction quality, and relief of ``cliff effect''. However, it is difficult to couple existing secure communication mechanisms (e.g., encryption-decryption mechanism) with DJSCC in contrast with traditional SSCC schemes, which hinders the practical usage of this emerging technology. To this end, our paper proposes a novel method called DL-based joint protection and source-channel coding (DJPSCC) for images that can successfully protect the visual content of the plain image without significantly sacrificing image reconstruction performance. The idea of the design is to use a neural network to conduct visual protection, which converts the plain image to a visually protected one with the consideration of its interaction with DJSCC. During the training stage, the proposed DJPSCC method learns: 1) deep neural networks for image protection and image deprotection, and 2) an effective DJSCC network for image transmission in the protected domain. Compared to existing source protection methods applied with DJSCC transmission, the DJPSCC method achieves much better reconstruction performance.

\end{abstract}


%
\IEEEpeerreviewmaketitle

\section{Introduction}
\label{Introduction}
\IEEEPARstart{T}{he} modular design principle based on Shannon's separation theorem \cite{cover1999elements} is the cornerstone of modern communications and has enjoyed great success in the development of wireless communications. However, the assumptions of unlimited codeword length, delay, and complexity in the separation theorem are not possible in real wireless environments, leading to sub-optimal separate source-channel coding (SSCC). Moreover, for time-varying channels, when the channel quality is worse than the target channel quality, SSCC cannot decode any information due to the collapse of channel coding; when the channel quality is better than the target quality, separate coding cannot further improve reconstruction quality. This is the famous ``cliff effect'' \cite{skoglund2006hybrid}, which increases the cost of SSCC during wireless transmission. In recent years, joint source-channel coding (JSCC) has been theoretically demonstrated to have better error exponents than SSCC in discrete memoryless source channels \cite{zhong2007joint}, which motivates the development of various JSCC designs over the years. Benefiting from the data-driven nature, deep learning (DL)-based JSCC (DJSCC) successfully reduces the difficulty of coding design existing in traditional JSCC for variant types of sources and channels \cite{bourtsoulatze2019deep, weng2021semantic, xu2022deep}, balances performance and storage requirements \cite{kurka2021bandwidth,kurka2020deepjscc,xu2022wireless}, cooperates with orthogonal frequency division multiplexing (OFDM) widely employed in wireless communication systems \cite{yang2022ofdm}, and matches semantic communications \cite{xu2022deepSematic}.

\begin{figure*}[t]
\centering
\includegraphics[width=2\columnwidth]{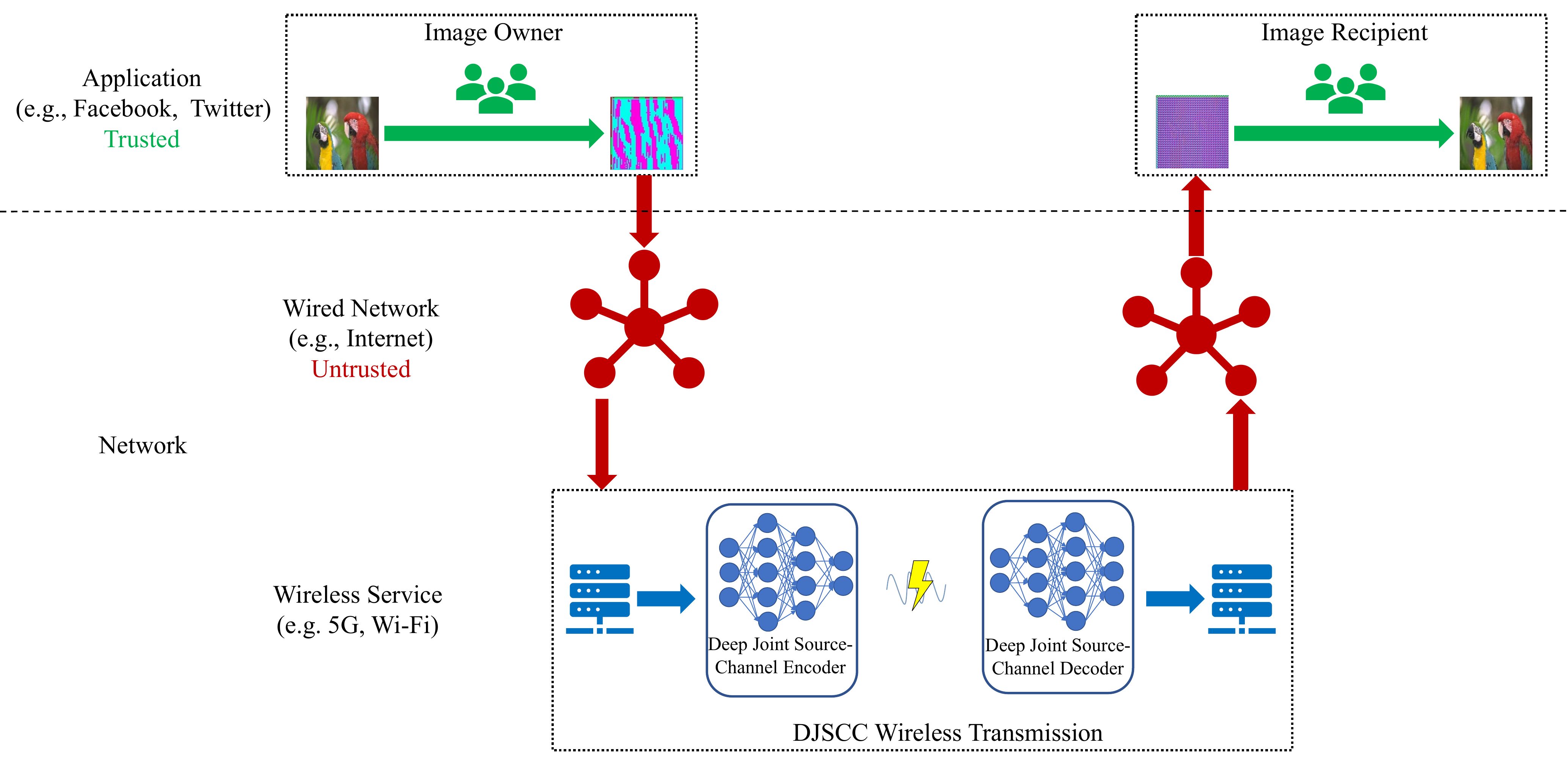}
\caption{The DJSCC based wireless communication system.}
\label{Fig:overview}
\end{figure*}

The important step to put DJSCC into practice is to protect the source information from eavesdroppers. Note that DJSCC is empowered by data-driven manner, which learns the effective encoding function and decoding function from scratch, causing the correlation between the source signal and the channel input symbols. To conduct a secure DJSCC communication, the source signal or the channel input symbols should be protected during DJSCC transmission. Considering that the channel input symbols belonging to the physical layer can be operated only by the wireless service provider in practical wireless scenarios, in this paper, we design to protect the source signal owned by the users, who can freely manipulate the source signal in the application layer. As illustrated in Fig.~\ref{Fig:overview}, the image owner intends to transmit a plain image to the image recipient through the network that contains an untrusted wired network and a wireless transmission service. To protect the visual content of the plain image, the image owner transforms the plain image into a protected image before providing it to the wireless service provider. Then the visually protected image is transmitted by the wireless service provider through DJSCC transmission. After DJSCC wireless transmission, the protected image decoded by the DJSCC decoder with some distortion is transmitted to the image recipient through the untrusted network (e.g., Internet). The image recipient transforms the distorted protected image to the plain image. Even if the protected image or the distorted protected image was leaked or stolen during the wired network transmission process, the visual content of the plain image cannot be acquired directly by eavesdroppers. 

One might like to use existing protection methods in SSCC-based communication systems, which can be conducted either before the source encoder or after the source encoder. However, a major issue with these protected methods of SSCC is that the change in the visual structure of the plain image leads to the degradation of the transmission of DJSCC. It is worth noting that the potential method applied in this scenario is not limited by an information-theoretic framework \cite{li2018fundamental}.

In this paper, we design a DL-based joint protection and source-channel coding method that can generate the visually protected image suitable for the DJSCC transmission. To the best of our knowledge, this is the first scheme to couple source protection with DJSCC based wireless transmission. The major contributions are summarized as follows.

\begin{itemize}
\item[$\bullet$]We design a unified framework which consists of the protection module, the deprotection module, the feature extraction module and the DJSCC module to protect the visual content of the image source taking into account DJSCC transmission, which can overcome the problem of ``cliff effect'', avoid the latency brought by channel mismatch, and most importantly, lead to considerable performance. 

\item[$\bullet$] By using the feature extraction module, we redesign the loss function of the proposed end-to-end framework. During the training stage, the proposed framework learns an effective method to provide visual protection for the plain image, an effective protection domain for the subsequent DJSCC transmission, an effective DJSCC transmission method, and an effective method to reconstruct the plain image. The strength of visual protection can be adjusted to satisfy different levels of protection requirements.

\item[$\bullet$] We propose two design principles for the protection module and the deprotection module. These principles can quickly guide the concrete design of the protection network and the deprotection network and meet the real communication scenario requirements, e.g., the storage overhead and computational complexity.

\end{itemize}

The rest of this paper is organized as follows. Section II presents related work on deep joint source-channel coding and image protection. Then, the proposed method is presented in Section III. In Section IV, the proposed method is evaluated on datasets with low resolution and high resolution. 
Finally, Section V concludes this paper.

\section{Related work}

\subsection{Deep Joint Source Channel Coding}

The initial DJSCC work proposed a recurrent neural network (RNN) for text transmission over binary erasure channels \cite{farsad2018deep}. From then on, DJSCC attracted increasing interest, especially for image compression and transmission. Compared to the SSCC scheme (e.g., JPEG/JPEG2000 for image coding and LDPC for channel coding), the DJSCC scheme designed in \cite{bourtsoulatze2019deep} has better image restoration quality, especially in the low signal-to-noise ratio (SNR) regime. To well adapt to variable bandwidth and deal with coding for distributed sources \cite{yeung1999distributed}, DJSCC schemes with adaptive-bandwidth image transmission and distributed transmission are proposed in \cite{kurka2021bandwidth} and \cite{wang2022distributed}, respectively. Taking into account the classical feedback scenario \cite{kostina2017joint}, image transmission with channel output feedback is proposed in \cite{kurka2020deepjscc}. However, all the aforementioned schemes are trained and deployed under the same channel conditions (the single SNR) to ensure optimality, demanding the use of multiple trained networks to suit a range of SNR that leads to considerable storage requirements in transceivers. To overcome this challenging problem in DJSCC, \cite{xu2022wireless} proposed a single network for DJSCC that can adapt to a wide range of SNR conditions to meet the memory limit of the device in real wireless scenarios. Furthermore, \cite{choi2019neural} proposed a DJSCC method based on the maximization of mutual information between the source and the received noisy codeword for the binary erasure channel and the binary symmetric channel. \cite{saidutta2019joint} and \cite{saidutta2019joint2} model their DJSCC systems via a variational autoencoder and manifold variational autoencoders for a Gaussian source, respectively. So far, by using the data-driven method, DJSCC successfully reduces the difficulty of coding design in traditional JSCC, making it a promising technology in low-latency and low-power scenarios.

\subsection{Image Protection}
\begin{figure}[t]
\centering
\subfigure[]{
\label{sscc_bfc}
\includegraphics[width=1\columnwidth]{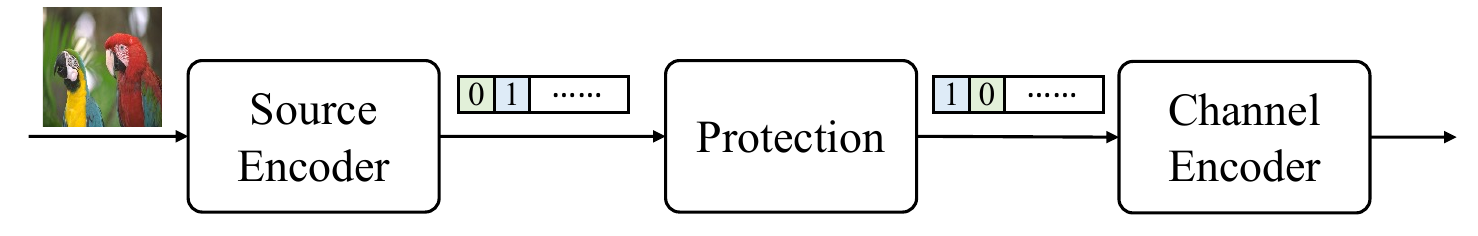}}
\subfigure[]{
\label{sscc_bfs}
\includegraphics[width=1\columnwidth]{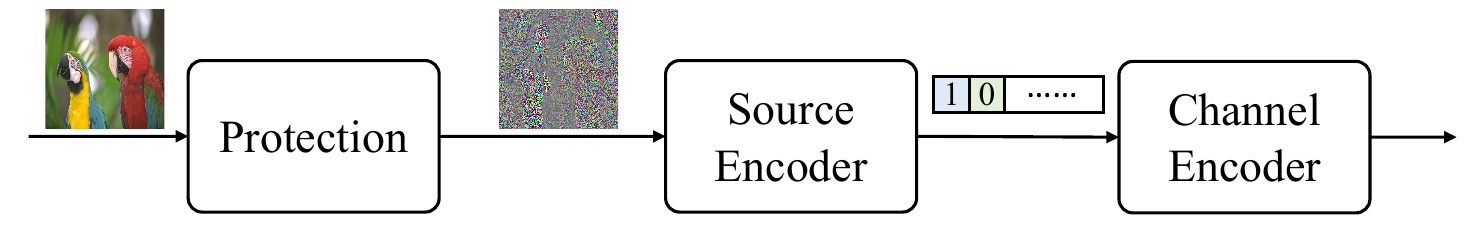}}
\subfigure[]{
\label{jscc_bfs}
\includegraphics[width=1\columnwidth]{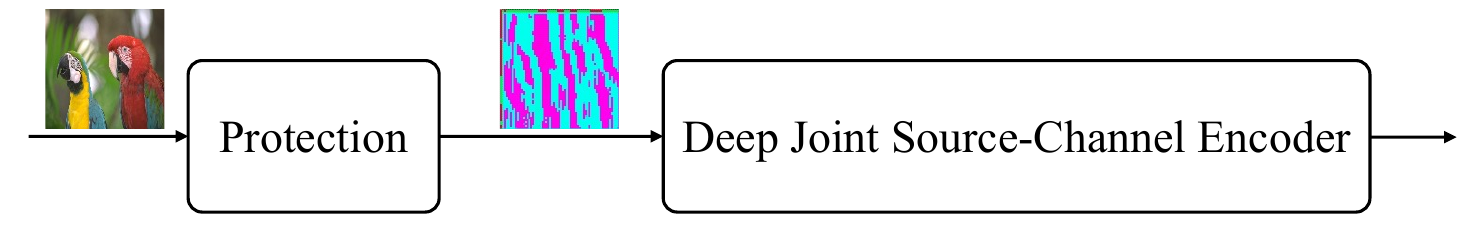}}
\caption{Different strategies for image protection in SSCC and DJSCC. (a) Protect the data encoded by the source encoder before the channel encoder in SSCC, (b) Protect the image source before the source encoder in SSCC, and (c) Protect the image source before the source encoder in DJSCC.
}\label{Fig:enc_loc}
\end{figure}

In SSCC, the protection operation can be executed between the source encoder and the channel encoder, as shown in Fig.~\ref{sscc_bfc}, or before the source encoder, as shown in Fig.~\ref{sscc_bfs}. The first strategy applies the source encoder to compress the image to the binary data, and then uses bit-oriented based encryption methods, e.g., the data encryption standard (DES) \cite{davis1978data}, advanced encryption standard (AES) \cite{heron2009advanced}, and Rivest–Shamir–Adleman (RSA) \cite{rivest1978method}, to protect the binary data. The second strategy is fit for the typical scenario, where the image provider only takes care of protecting the image content and the telecommunications provider has an overriding interest in improving spectrum efficiency. To directly protect the image content, various pixel-based and space-based image protection methods have been developed to transform a plain image to a visually protected image, e.g., Arnold \cite{wu2009arnold}, Hill algorithm \cite{acharya2009image} and 3D chaotic map-based method \cite{kanso2012novel}. Taking into account the compression needs in the second strategy, dedicated protection and compression methods are proposed to improve the compression ratio in the protection domain\cite{zhou2014designing, kang2013performing, chuman2019encryption}. As shown in Fig.~\ref{jscc_bfs}, to protect the visual content of the plain image for DJSCC transmission, the image protection should be executed in front of the deep joint source-channel encoder, which is similar to the position of the protection module in Fig. 2(b). The protection methods originally designed for the second strategy in SSCC may be applied in DJSCC. However, these image protection methods change the visual structure of the plain image, break the coherence in adjacent pixels, and may cause performance degradation of DJSCC transmission.

DL has led to state-of-the-art performance in various image processing tasks, motivating the application of DL to protect the image source. Image protection methods proposed by \cite{ito2021image,sirichotedumrong2021gan} are designed for DL-based classification with acceptable classification accuracy. However, these image protection methods designed for the classification task are not suitable for DJSCC. Protected images for the classification task only reserve some specific semantic information relevant to the image class, while the pixel-based information of the image is discarded, which causes the performance degradation for the image reconstruction task.

\section{Deep Joint Protection and Source-Channel Coding}
Based on the scenario described in Section \ref{Introduction}, our motivation is to successfully protect the visual content of the plain image without significantly sacrificing the image reconstruction performance and efficiently transmit the visually protected image through the DJSCC. In this Section, a DL-based joint protection and source-channel coding (DJPSCC) method is proposed for these purposes. 

\subsection{System Model}
\begin{figure*}[!tb]
\centering
\includegraphics[width=2\columnwidth]{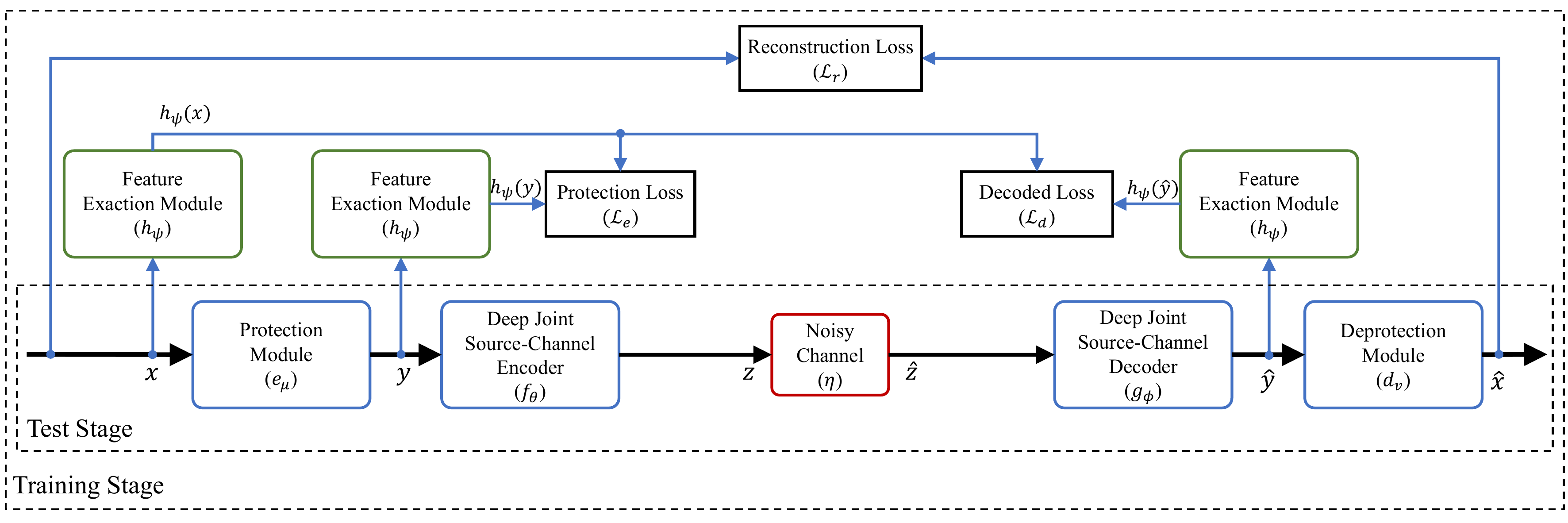}
\caption{The system model of the proposed DJPSCC method.}\label{Fig:edjscc}
\end{figure*}

Considering a visually protected DJSCC transmission system, as shown in the lower part of Fig.~\ref{Fig:edjscc}, a plain image is represented by $\boldsymbol{x} \in \mathbb{R}^n$, where $\mathbb{R}$ denotes the set of real numbers and $n=h \times w \times c$. Here, $h, w$ and $c$ denote the height, the width and the number of channels of an image, respectively. The protection module transforms the plain image into a visually protected image, which is expressed as:
\begin{equation} 
\boldsymbol{y}=e_{\boldsymbol{\mu}}(\boldsymbol{x}) \in \mathbb{R}^n,
\label{encrypt}
\end{equation}
where $e_{\boldsymbol{\mu}}(\cdot)$ represents an protection module parameterized by the set of parameters $\boldsymbol{\mu}$. The protected image $\boldsymbol{y}$ and the plain image $\boldsymbol{x}$ have the same size. 

After the protection process, the protected image $\boldsymbol{y}$ is encoded by the joint source-channel encoder as:
\begin{equation} 
\boldsymbol{z}=f_{\boldsymbol{\theta}}(\boldsymbol{y}) \in \mathbb{C}^k,
\label{encode}
\end{equation}
where $\mathbb{C}$ denotes the set of complex numbers, $k$ represents the number of channel input symbols, and $f_{\boldsymbol{\theta}} (\cdot)$ represents a joint source-channel encoder parameterized by the set of parameters $\boldsymbol{\theta}$. The real and imaginary parts of $\boldsymbol{z}$ are mapped to the in-phase components I and the quadrature components Q of the transmitted signals, respectively. During transmission, the power constraint $ \frac{1}{k}\mathbb{E}(\boldsymbol{zz^*})\leq 1$ must be satisfied, where $\boldsymbol{z^*}$ is the complex conjugate transpose of $\boldsymbol{z}$. 

The transmitted signals are corrupted by the wireless channel. We adopt the well known additive white Gaussian noise (AWGN) model given by:
\begin{equation} 
\boldsymbol{\hat{z}}=\eta(\boldsymbol{z})= \boldsymbol{z}+ \boldsymbol{\omega}, 
\label{awgn}
\end{equation}
where $\boldsymbol{\hat{z}} \in \mathbb{C}^k$ is the channel output and $\boldsymbol{\omega} \in \mathbb{C}^k$ denotes the additive noise modeled by $\boldsymbol{\omega} \sim \mathbb{CN}(0, \sigma^2\boldsymbol{I})$, where $\sigma^2$ represents the average noise power and $\mathbb{CN}(\cdot,\cdot)$ denotes a circularly symmetric complex Gaussian distribution.

In turn, the channel output symbols $\boldsymbol{\hat{z}}$ are decoded by the joint source-channel decoder as:
\begin{equation} 
\boldsymbol{\hat{y}}=g_{\boldsymbol{\phi}}(\boldsymbol{\hat{z}}) \in \mathbb{R}^n, 
\label{decode}
\end{equation}
where $g_{\boldsymbol{\phi}} (\cdot)$ represents a joint source-channel decoder parameterized by the set of parameters $\boldsymbol{\phi}$. The decoded image $\boldsymbol{\hat{y}} \in \mathbb{R}^n$ should be a visually protected image, with the same size as the protected image $\boldsymbol{y}$.

Similarly to the protection step, the deprotection module is employed to convert the decoded image to the deprotected image as follows:
\begin{equation}
\boldsymbol{\hat{x}}=d_{\boldsymbol{\nu}}(\boldsymbol{\hat{y}}) \in \mathbb{R}^n, 
\label{decrypt}
\end{equation}
where $d_{\boldsymbol{\nu}} (\cdot)$ represents a deprotection module parameterized by the set of parameters $\boldsymbol{\nu}$ and the deprotected image $\boldsymbol{\hat{x}} \in \mathbb{R}^n$ is a restored version of the plain image. The bandwidth ratio $R$ is defined as $k/n$, where $n$ is the source size (i.e., image size) and $k$ is the channel bandwidth (i.e., channel input size). 

\subsection{The Proposed Method}
In sharp contrast to DJSCC methods \cite{bourtsoulatze2019deep, kurka2020deepjscc, kurka2021bandwidth, xu2022wireless}, we require our DJPSCC method to address two issues:

1) Protect the visual content of a plain image.

2) Extract effective features from the protected image for subsequent DJSCC transmission.

\begin{figure}[!tb]
\centering
\includegraphics[width=1\linewidth]{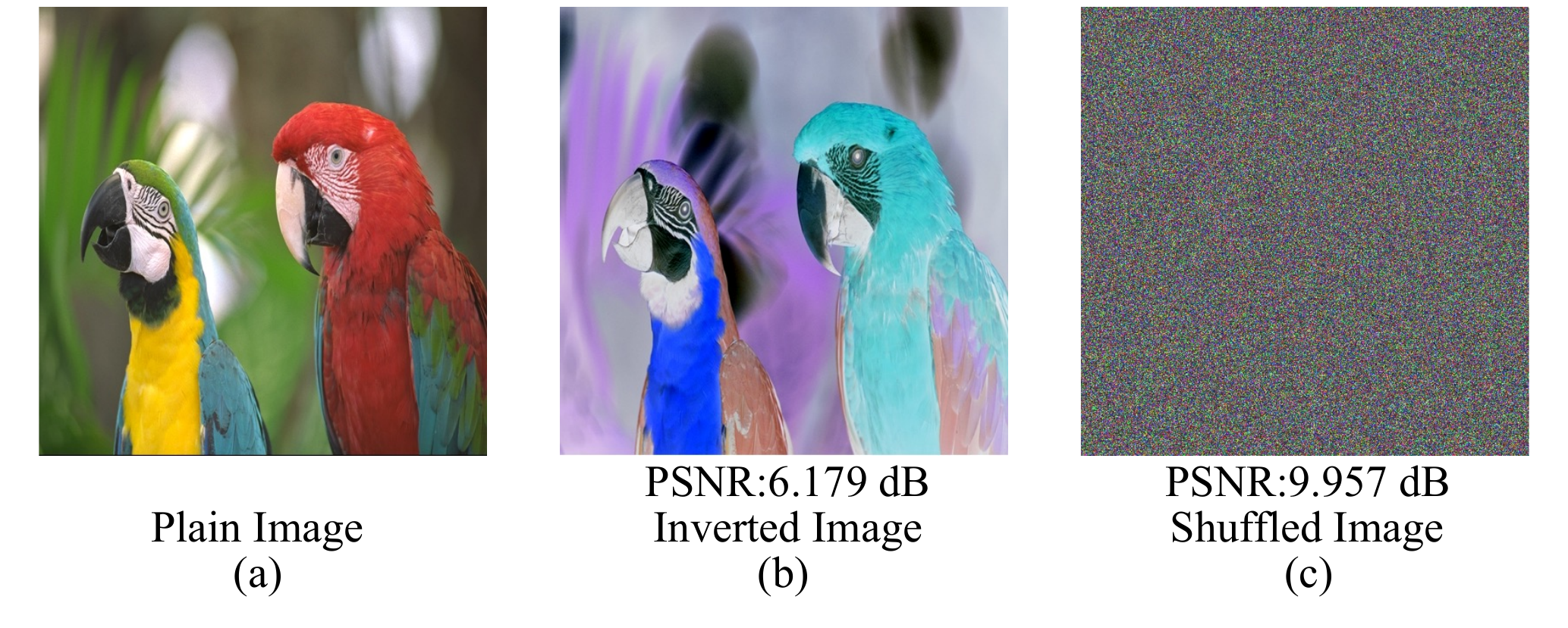}
\caption{PSNR comparison. (a) plain image, (b) inverted image (the intensity values of the plain image are subtracted by 255), (c) shuffled image (the intensity values of the plain image are randomly shuffled in space and channel dimension).}\label{Fig:psnr_shortage}
\end{figure}

The classical full-reference metric of image similarity is peak signal-to-noise ratio (PSNR) between the original image and the restored image, which is defined as:
\begin{equation} 
\rm PSNR = 10log_{10}\frac{MAX^2}{MSE}(dB). 
\label{psnr}
\end{equation}
where MAX is the maximum possible value of the image pixels and MSE is the abbreviation of mean square error between the original image and the restored image. Although the prediction of PSNR performance is not always consistent with visual quality perceived by the human visual system, its simplicity makes it widely used in the field of image processing \cite{mohammadi2014subjective}. However, as illustrated in Fig.~\ref{Fig:psnr_shortage}, PSNR is not a good metric to assess visual security due to the excessive difference between the plain image and the visually protected image. An example is that the image in Fig.~\ref{Fig:psnr_shortage}(b) has a lower PSNR than that in Fig.~\ref{Fig:psnr_shortage}(c), while the visual content (e.g., the birds and the leaf) in Fig.~\ref{Fig:psnr_shortage}(b) are more easily identified than the image in Fig.~\ref{Fig:psnr_shortage}(c). 

In recent years, various visual security metrics (VSMs), including handcraft-based VSMs \cite{mao2006joint, tong2010visual,xiang2020visual,wen2021visual} and DL-based VSMs \cite{yue2019no, yang2021convolutional}, are designed to assess the visual security of the image. Here, we employ a feature extraction network to measure visual security. The feature extraction method has been used successfully to measure the similarity between two images \cite{johnson2016perceptual} and the difference between two images \cite{ito2021image}.

Concretely, in the training stage, the features of the plain image $\boldsymbol{x}$, the protected image $\boldsymbol{y}$, and the decoded image $\boldsymbol{\hat{y}}$ are extracted by the feature extraction module $h_{\boldsymbol{\psi}}$ in Fig.~\ref{Fig:edjscc}, where $\psi$ is the set of parameters of the feature extraction module. Note that both the protected image $\boldsymbol{y}$ and the decoded image $\boldsymbol{\hat{y}}$ are in the protected domain. The feature loss $\mathcal{L}_e$ between the plain image $\boldsymbol{x}$ and the protected image $\boldsymbol{y}$ is expressed as:
\begin{equation} 
\mathcal{L}_e = \frac{1}{m}\lVert h_{\boldsymbol{\psi}}(\boldsymbol{x})-h_{\boldsymbol{\psi}}(\boldsymbol{y}) \rVert_2^2, 
\label{f_loss_e}
\end{equation}
and the feature loss $\mathcal{L}_d$ between the plain image $\boldsymbol{x}$ and the decoded image $\boldsymbol{\hat{y}}$ is expressed as:
\begin{equation} 
\mathcal{L}_d = \frac{1}{m}\lVert h_{\boldsymbol{\psi}}(\boldsymbol{x})-h_{\boldsymbol{\psi}}(\boldsymbol{\hat{y}}) \rVert_2^2, 
\label{f_loss_d}
\end{equation}
where $h_{\boldsymbol{\psi}}(\boldsymbol{x}) \in \mathbb{R}^{m}$, $h_{\boldsymbol{\psi}}(\boldsymbol{y}) \in \mathbb{R}^{m}$, and $h_{\boldsymbol{\psi}}(\boldsymbol{\hat{y}}) \in \mathbb{R}^{m}$ are the features of the plain image $\boldsymbol{x}$, the protected image $\boldsymbol{y}$, and the decoded image $\boldsymbol{\hat{y}}$ extracted by the feature extraction network and $m=h_f \times w_f \times c_f$. Here, $h_f$, $w_f$, and $c_f$ denote the height, width, and number of channels of an extracted feature, respectively. For simplicity, MSE is adopted in this paper to characterize the strength of visual security. Other forms of the feature loss can also be applied in the proposed DJPSCC method. However, since we focus on the mechanism design, the design of the feature loss is beyond the scope of this work. It is worth noting that once the feature extraction module is chosen, its parameters are fixed during the training stage. 

Moreover, the reconstruction loss $\mathcal{L}_{r}$ between the plain image $\boldsymbol{x}$ and the deprotected image $\boldsymbol{\hat{x}}$ is expressed as:
\begin{equation} 
\mathcal{L}_{r}=d(\boldsymbol{x},\boldsymbol{\hat{x}})=\frac{1}{n} \lVert \boldsymbol{x}- \boldsymbol{\hat{x}} \rVert_2^2 =\frac{1}{n}\sum_{i=1}^n(x_i-\hat{x}_i)^2,
\label{single_distortion}
\end{equation}
where $x_i$ and $\hat{x}_i$ represent the $i$-th pixel value of the plain image $\boldsymbol{x}$ and the deprotected image $\boldsymbol{\hat{x}}$, respectively.

Unlike image-to-image translation tasks \cite{sirichotedumrong2021gan, johnson2016perceptual} that minimize feature loss in the training stage, the proposed method maximizes $\mathcal{L}_e$ and $\mathcal{L}_d$ to provide visual protection in protected images $\boldsymbol{y}$ and $\boldsymbol{\hat{y}}$. The total loss applied to train the proposed DJPSCC method is expressed as:
\begin{equation} 
\mathcal{L}_{total} = \mathcal{L}_{r}-\lambda_e\mathcal{L}_e-\lambda_d\mathcal{L}_d, 
\end{equation}
where $\lambda_e \in \mathbb{R}^+$ and $\lambda_d \in \mathbb{R}^+$ are the weights of $\mathcal{L}_e$ and $\mathcal{L}_d$, respectively. It should be emphasized that the feature loss $\mathcal{L}_e$/$\mathcal{L}_d$ represents the difference between the plain image and the protected/decoded image. A smaller value of $\mathcal{L}_e$/$\mathcal{L}_d$ means a small visual difference between the plain image and the protected/decoded image, while a larger value means a big difference between the plain image and the protected/decoded image and is considered to protect the visual content of the plain image. During the training stage, when the total loss $\mathcal{L}_{total}$ is minimized, its component $\mathcal{L}_e$/$\mathcal{L}_d$ loss is maximized to increase the visual protection ability.

Under a certain bandwidth ratio $R=k/n$, the proposed DJPSCC method learns the parameters of the protection module $\boldsymbol{\mu}$, the deep joint source-channel encoder $\boldsymbol{\theta}$, the joint source-channel decoder $\boldsymbol{\phi}$, and the deprotection module $\boldsymbol{\nu}$ by minimizing the total loss as follows:
\begin{equation} 
(\boldsymbol{\mu}^*,\boldsymbol{\theta}^*, \boldsymbol{\phi}^*,\boldsymbol{\nu}^*)=\mathop{\arg\min}\limits_{\boldsymbol{\mu}, \boldsymbol{\theta},\boldsymbol{\phi}, \boldsymbol{\nu}}\mathbb{E}_{p(\sigma^2)}\mathbb{E}_{p(\boldsymbol{x},\boldsymbol{\hat{x}})}(L_{total}),
\label{argmin}
\end{equation}
where $\boldsymbol{\mu}^*,\boldsymbol{\theta}^*, \boldsymbol{\phi}^*,\boldsymbol{\nu}^*$ are the optimal parameters, $p(\boldsymbol{x},\boldsymbol{\hat{x}})$ represents the joint probability distribution of the plain image $\boldsymbol{x}$ and the deprotected image $\boldsymbol{\hat{x}}$, $\sigma^2$ is the average noise power, and $p(\sigma^2)$ represents the probability distribution of the channel noise. Note that the probability distribution of the channel noise instead of the fixed channel noise is adopted in this paper with consideration of the storage overhead and the difficulty to acquire the signal-to-noise ratio (SNR) in the image owner/recipient. In addition, an empirical average instead of a statistical average is adopted in the training stage. During the training stage, the proposed DJPSCC method learns: 1) an effective method to provide visual protection for the plain image, 2) an easy-to-be-extracted image domain for the subsequent DJSCC transmission, 3) an effective DJSCC transmission method, and 4) an effective method to reconstruct the plain image.

After training the DJPSCC network, to provide security guarantees against the eavesdropper, the protection module $e_{\boldsymbol{\mu}}$ and the deprotection module $d_{\boldsymbol{\nu}}$ are distributed securely to the owner of the image and the recipient of the image using the security protocol, e.g., the Secure Sockets Layer (SSL) protocol, respectively. The deep joint source-channel encoder and decoder are distributed to the DJSCC transmission service provider. However, the protected/decoded image is correlated with the plain image because of the inherent transform operation existing in the protection stage, causing a potential weakness when attacked by applying a generative adversarial network with abundant data (e.g., protected images and irrelevantly plain images) and sufficient computational resource. To enhance the security of the proposed method, multiple DJPSCC networks could be trained with different initialized parameters, and the network itself could be the key to secure transmission.

In the test stage, the plain image $\boldsymbol{x}$ is first converted to the protected image $\boldsymbol{y}$ by the image owner using Eq. \eqref{encrypt}. The protected image $\boldsymbol{y}$ is then sent to the DJSCC transmission service provider. DJSCC transmission is executed using Eq. \eqref{encode} and Eq. \eqref{decode} and the decoded image $\boldsymbol{\hat{y}}$ is obtained. The DJSCC transmission service provider sends the decoded image $\boldsymbol{\hat{y}}$ to the image recipient, which uses Eq. \eqref{decrypt} to deprotected the decoded image $\boldsymbol{\hat{y}}$. The test stage process is illustrated in the lower part of Fig.~\ref{Fig:edjscc}. 

\section{Experimental Results}
\label{Experimental Results}

\begin{figure}[!tb]
\centering
\includegraphics[width=1\columnwidth]{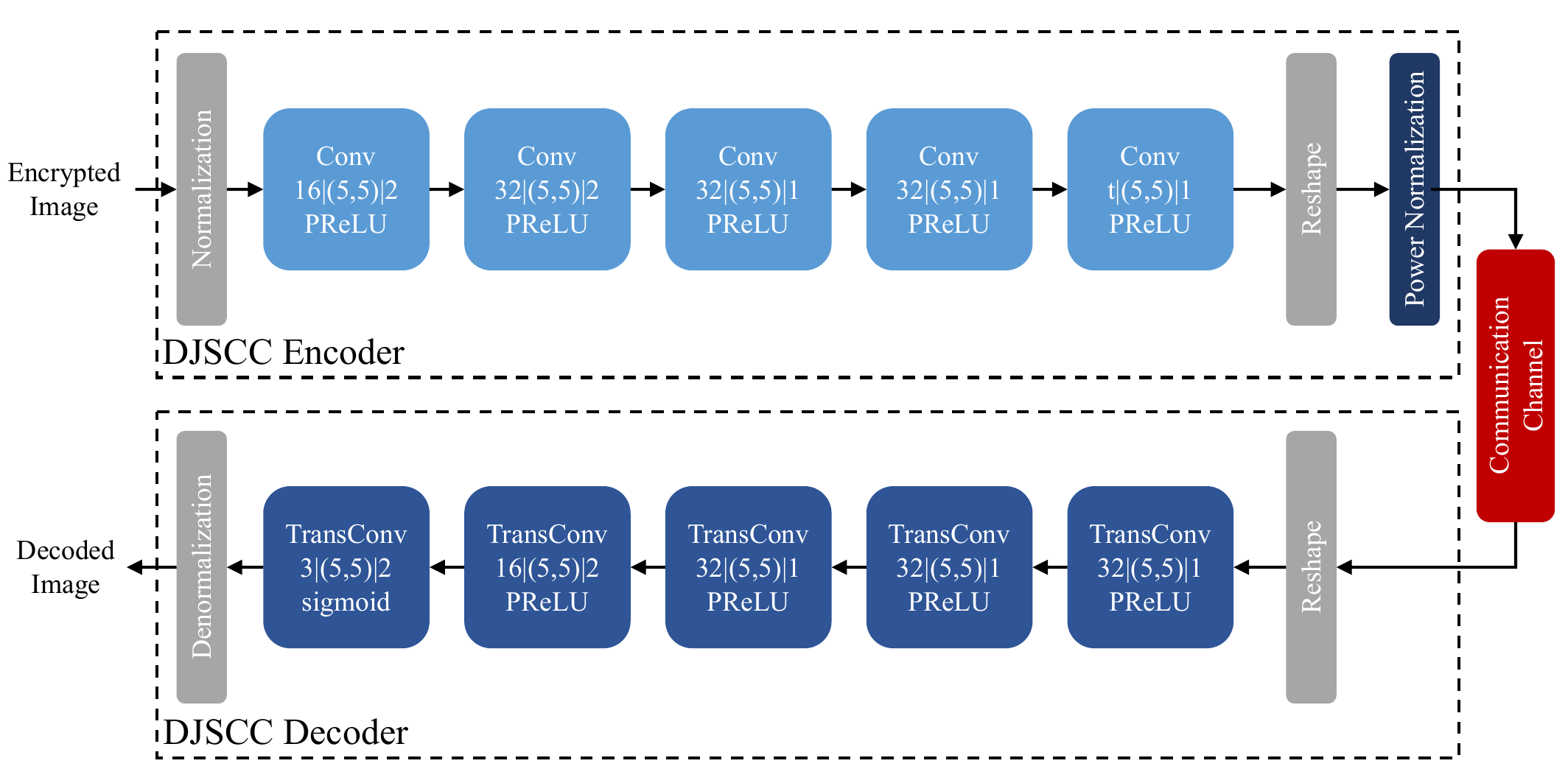}
\caption{The architecture of the DJSCC network\cite{bourtsoulatze2019deep} adopted in this paper.  The notation $K|(F, F)|S$ in a convolutional/transposed convolutional layer denotes that it has $K$ filters with size $F$ and stride down/up $S$.}\label{Fig:djscc_arch}
\end{figure}

The proposed DJPSCC is a general framework that can be employed in most existing DJSCC architectures. To demonstrate the effectiveness of the DJPSCC, the first DJSCC architecture proposed in \cite{bourtsoulatze2019deep} is adopted in subsequent experiments. As shown in Fig.~\ref{Fig:djscc_arch}, the DJSCC encoder consists of the normalization layer, five alternant convolutional layers and PReLU layers, the reshape layer, and the power normalization layer. The DJSCC decoder consists of the reshape layer, five alternant transposed convolutional layers and activation layers (i.e., four PReLU layers and one sigmoid layer), and the denormalization layer. The normalization layer converts the input image with the pixel value range [0, 255] to the image with the pixel value range [0, 1], and the denormalization layer performs the opposite operation. The notation $K|(F, F)|S$ in a convolutional/transposed convolutional layer denotes that it has $K$ filters with size $F$ and stride down/up $S$. The power normalization layer is used to satisfy the average power constraint at the transmitter. The channel number of the last convolutional layer in the DJSCC encoder is $t$, which is relevant to the channel bandwidth.

Although the power of the proposed DJPSCC is based on the ingenious design of the loss function and the end-to-end training strategy, the architectures of the protection module and the deprotection module still affect the reconstruction performance of the proposed DJPSCC. Here, we adopt two principles in designing the protection/deprotection module: 1) Shortcut connections in the protection/deprotection network could enhance the reconstruction performance of the proposed DJPSCC, while 2) A deeper protection/deprotection network would degrade the reconstruction performance of the proposed DJPSCC. The effectiveness of the two principles will be demonstrated in \ref{ablation}.

Tensorflow \cite{abadi2016tensorflow} and its high-level API Keras are used to implement the proposed DJPSCC method. 
DJPSCC is trained with a uniform distribution within the SNR range [0, 20] dB. The following experiments are run on a Linux server with twelve octa-core Intel(R) Xeon(R) Silver 4110 CPUs and sixteen GTX 1080Ti GPUs. Each experiment was assigned six CPU cores and a GPU. 

\subsection{DJPSCC Validity on the CIFAR-10 Dataset}
\label{validity}

\begin{figure}[!tb]
\centering
\includegraphics[width=1\columnwidth]{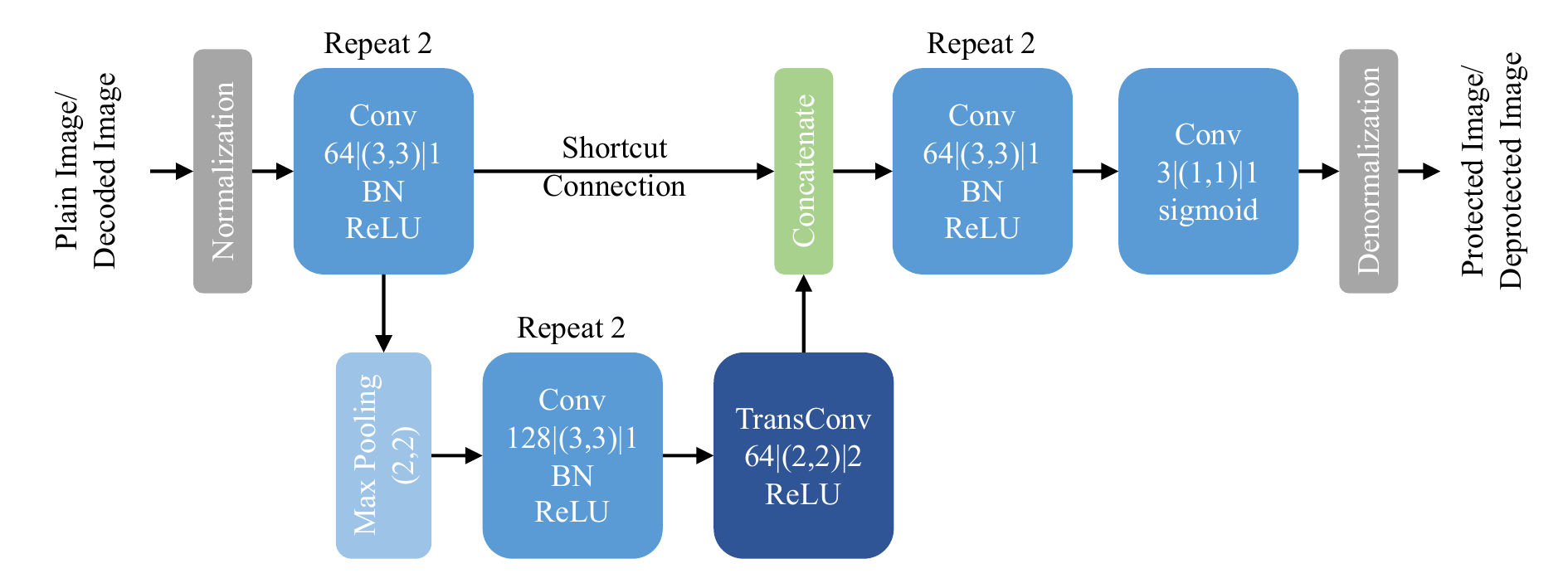}
\caption{The shallow version of U-Net \cite{ronneberger2015u} with one shortcut connection (UNet-S1) is adopted as the protection/deprotection network in this paper.}\label{Fig:trans_arch}
\end{figure}
We first consider the performance of the proposed method on the CIFAR-10 dataset, which consists of 60000 $32 \times 32\times3$ color images associated with 10 classes where each class has 6000 images. Note that the goal of our proposed method is to generate visually protected images for the untrusted transmission channels and reconstruct the plain image at the receiver, so the class label of each image is useless in the following experiments. The training dataset and the test dataset contain 50000 images and 10000 images, respectively. 

U-Net \cite{ronneberger2015u} owns shortcut connections from its contracting path to its expansive path since it meets our first principle for the protection/deprotection module. As shown in Fig.~\ref{Fig:trans_arch}, a shallow version of U-Net with one shortcut connection (UNet-S1) is designed as the protection/deprotection network. The VGG16\footnote{https://keras.io/api/applications/vgg/\#vgg16-function} \cite{simonyan2014very} pretrained on ImageNet dataset is adopted as the feature extraction network. All networks were trained for 500 epochs using Adam Optimizer with an initial learning rate of $10^{-3}$. Once learning stagnated for 10 epochs, the learning rate was reduced by a factor of 10.  The performance of the DJPSCC networks was evaluated at specific $\rm SNR_{test} \in$ [0,20] dB on the CIFAR-10 test dataset. To alleviate the effect of the randomness caused by the wireless channel, each image in the CIFAR-10 test dataset is transmitted 10 times. PSNR is used in the evaluation of the reconstruction performance between the plain image and the deprotected image. For simplicity, we allocate the same loss weight for visually protected images as $\lambda_e=\lambda_d=\lambda$.

\begin{figure}[!tb]
\centering
\includegraphics[width=1\columnwidth]{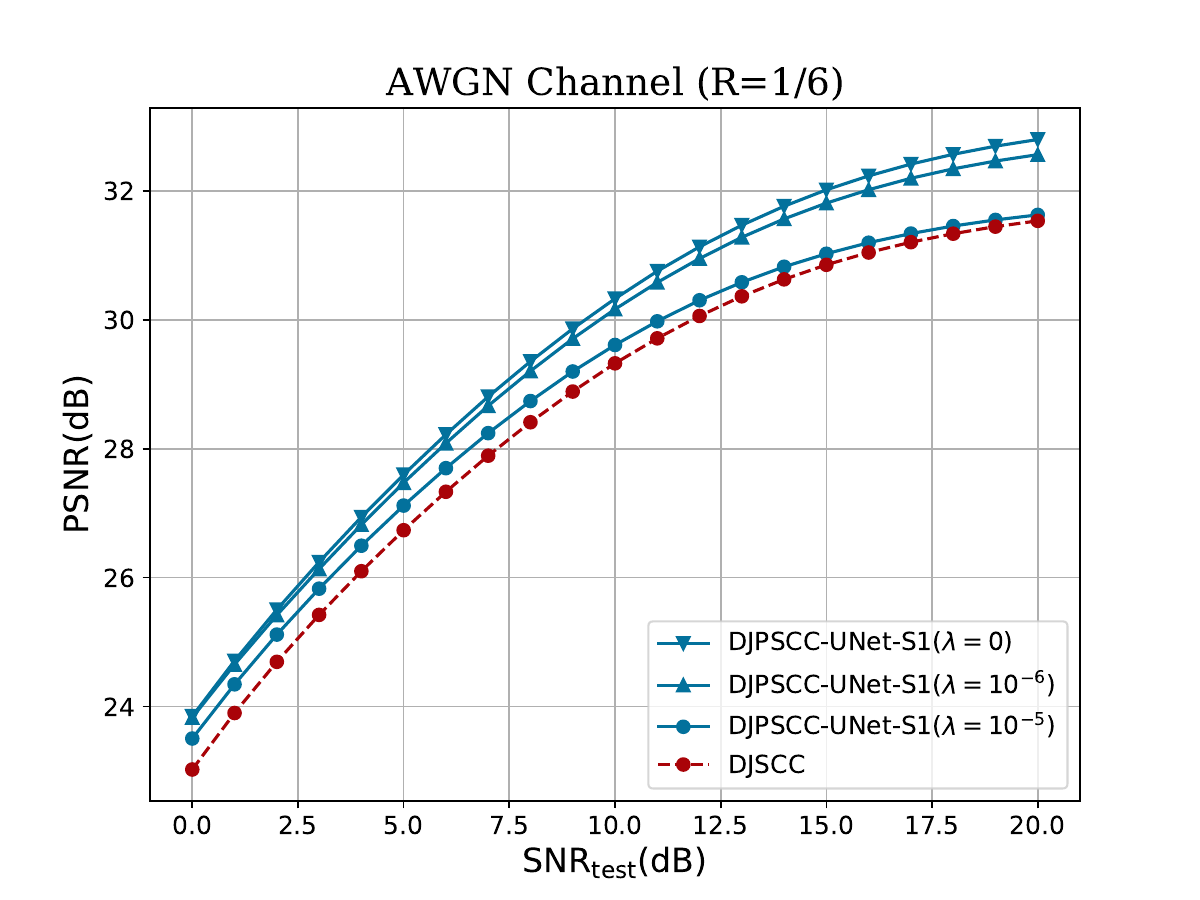}
\caption{Reconstruction performance of DJPSCC-UNet-S1 and DJSCC on CIFAR-10 test dataset with bandwidth ratio R=1/6.}
\label{Fig:cifar10_c16_snrdb0to20}
\end{figure}

\begin{table*}[!tb] 
\renewcommand{\arraystretch}{1.3} \scriptsize
\caption{Visual Security Evaluation of DJPSCC-UNet-S1 on the CIFAR-10 test dataset} 
\label{Table:vse_unet}
\centering
\begin{tabular}{|c|cccc|cccc|cccc|}
\hline
\multirow{2}*{\textbf{Method}} & \multicolumn{4}{c|}{\textbf{Protected Image}}& \multicolumn{4}{c|}{\textbf{Decoded Image}}&\multicolumn{4}{c|}{\textbf{Deprotected Image}}\\\cline{2-13}
~&\textbf{LFBVS}&\textbf{COR}&\textbf{PSNR(dB)}&\textbf{SSIM}&\textbf{LFBVS}&\textbf{COR}&\textbf{PSNR(dB)}&\textbf{SSIM}&\textbf{LFBVS}&\textbf{COR}&\textbf{PSNR(dB)}&\textbf{SSIM}
\\\hline
DJPSCC-UNet-S1 ($\lambda=0$)    & 0.529 & 0.085 & 10.034 & 0.113  & 0.604 & -0.012 & 8.091 & -0.004 & \textbf{0.198} & \textbf{0.984} & \textbf{30.330} & \textbf{0.947} \\\hline
DJPSCC-UNet-S1 ($\lambda=10^{-6}$)& 0.684 & -0.006 & 6.280 & 0.004  & 0.674 & \textbf{-0.058} & 6.897 & \textbf{-0.019} & 0.202 & 0.983 & 30.162 & 0.944 \\\hline
DJPSCC-UNet-S1 ($\lambda=10^{-5}$) & \textbf{0.721} & \textbf{-0.082} & \textbf{4.980} & \textbf{-0.031}  & \textbf{0.675} & -0.010 & \textbf{5.420} & 0.001  & 0.209 & 0.981 & 29.686 & 0.939 \\\hline
\end{tabular}
\end{table*}

Fig.~\ref{Fig:cifar10_c16_snrdb0to20} compares the reconstruction performance of DJPSCC-UNet-S1 with different loss weights (e.g., $\lambda=0,10^{-6},10^{-5}$) at bandwidth ratio $R=1/6$. The reconstruction performance of the DJSCC without the protection module and the deprotection module is also plotted as a reference. With increasing $\lambda$, DJPSCC pays more attention to {visual protection tasks,} which cause a {degradation in reconstruction performance}. Since DJPSCC-UNet-S1 network is deeper and with more parameters than the DJSCC network, the reconstruction performance of DJPSCC-UNet-S1($\lambda=0$) without feature loss constraint is almost 2 dB better than that of the DJSCC in $\rm SNR_{test} \in [0,20]$ dB. Although the feature loss is imposed on DJPSCC, benefited from the powerful ability of DJPSCC-UNet-S1, the reconstruction performance of DJPSCC-UNet-S1($\lambda=10^{-5}$) is still better than that of DJSCC. 

Table \ref{Table:vse_unet} evaluates the visual security of DJPSCC-UNet-S1 using the VSMs, i.e., LFBVS \cite{tong2010visual} and correlation (COR) \cite{mohammad2017survey}. Note that the PSNR and the structural similarity index (SSIM) are frequently used as reference in visual security evaluation, so the PSNR and the SSIM are also listed in Table \ref{Table:vse_unet}. During the evaluation, $\rm SNR_{test}$ is 10 dB. A high score of the COR, the PSNR and the SSIM reflects high similarity between the visually protected image and the plain image, while a high score of the LFBVS reflects high visual security when comparing the visually protected image with the plain image. The range of PSNR, SSIM and LFBVS is $[0, +\infty]$, $(-1, 1]$ and $[0,1]$, respectively. With increasing $\lambda$, the PSNR and SSIM of the protected image gradually decrease and the LFBVS gradually increases. Evaluation of the decoded image with increasing $\lambda$ reveals a similar trend, except for a small inconsistency in the COR metric and the SSIM metric. Although the SSIM and the PSNR can exhibit satisfactory performance in predicting image quality, they are not appropriate for evaluating the visual security of protected images. For example, SSIM failed to measure badly blurred images \cite{chen2006gradient}. In this situation, the LFBVS metric is more accurate than the SSIM metric, since the LFBVS is specially designed to measure the visual security between two images. In the following experiments, DJPSCC-UNet-S1($\lambda=10^{-5}$) is chosen for the subsequent comparison.

\begin{figure}[!tb]
\centering
\includegraphics[width=0.8\columnwidth]{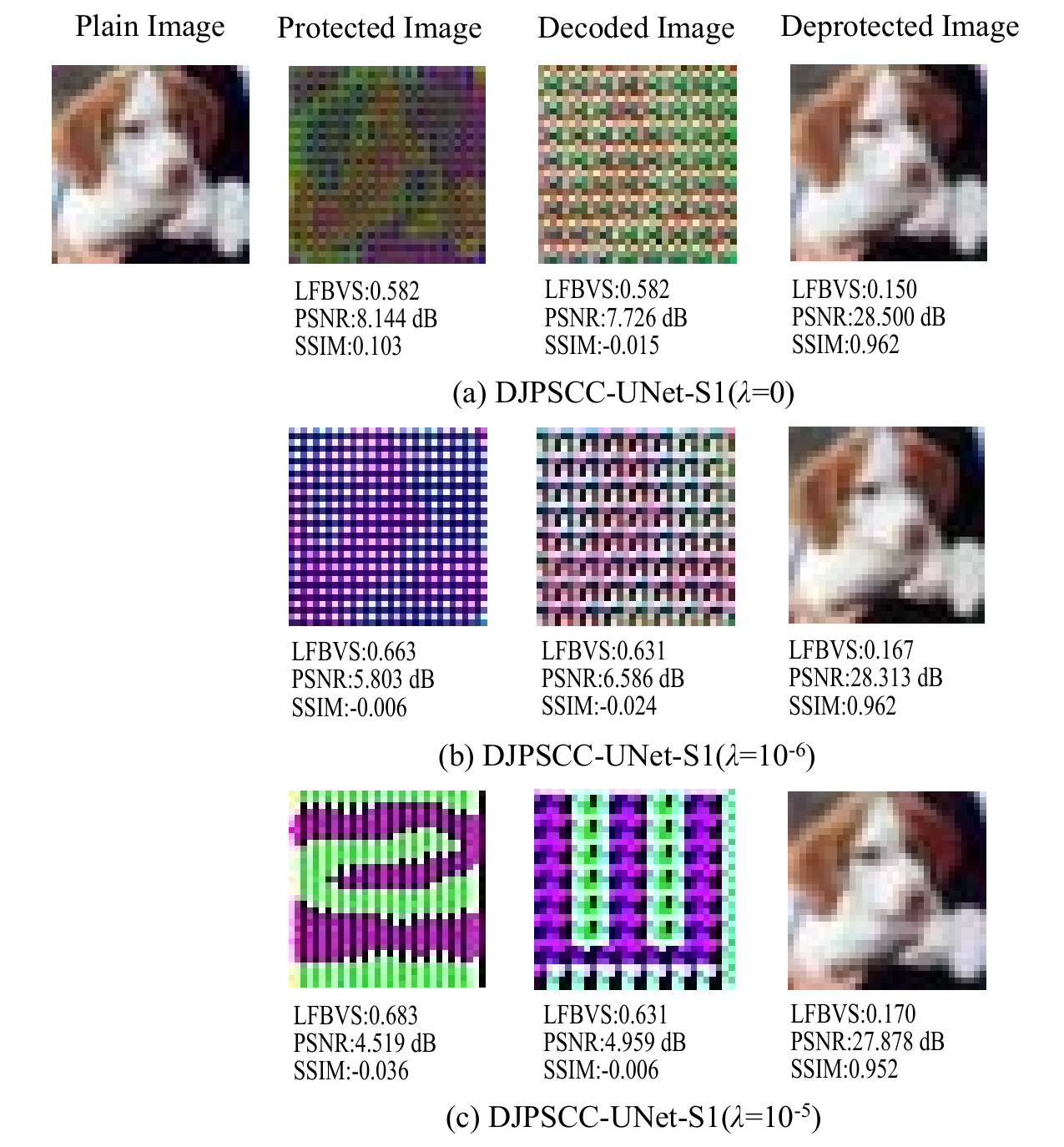}
\caption{Visually protected images and reconstruction images generated by DJPSCC-UNet-S1. The image in the first column is the plain image. The images in the second column are the protected images transformed by the image owner. The images in the third column are the decoded images decoded by the DJSCC decoder at $\rm SNR=10$ dB. The images in the last column are the deprotected images transformed by the image recipient. (a) $\lambda=0$, (b) $\lambda=10^{-6}$, (c) $\lambda=10^{-5}$.}\label{Fig:cifar10_c16_visual}
\end{figure}

Fig.~\ref{Fig:cifar10_c16_visual} shows the visualization of the plain image, the protected images transformed by the owner of the image, the decoded images decoded by the DJSCC transmission service provider, and the deprotected images transformed by the image recipient at $\rm SNR=10$ dB with different loss weights. The plain image comes from the CIFAR-10 test dataset. The outline of the dog can be vaguely identified in the protected image and the decoded image when $\lambda=0$. With increasing $\lambda$, the outline of the dog gradually disappears in the protected image and the decoded image. The most successful visual protection exists in the decoded image with $\lambda=10^{-5}$. All deprotected images with different $\lambda$ can reconstruct the main visual content conveyed by the plain image. Although there is no explicit visual security in the loss function when $\lambda=0$, the visual content is weakly protected to some extent due to the transformation provided by the protection module. If strong visual security is needed, a trade-off between the reconstruction performance and the visual protection performance exists in the DJPSCC method. That is, with increasing $\lambda$, the visual protection ability of the DJPSCC increases, while the quality of the DJPSCC reconstruction decreases.

This is the first work to design the DJPSCC framework that can protect the visual content of the plain image for DJSCC transmission. Here, we compare DJPSCC with one SSCC-based visual protection method, i.e., encryption then compression (EtC)\cite{chuman2019encryption}, and two visual protection methods designed for DL-based classification task, i.e., the learnable image encryption (LE) method \cite{tanaka2018learnable} and the pixel-based image encryption (PE) method \cite{sirichotedumrong2019privacy}. The EtC method is a block scrambling-based encryption scheme with JPEG compression, which can securely transmit plain images through an untrusted channel, e.g., the untrusted wired network in Fig.~\ref{Fig:overview}. In the SSCC-based wireless communication system, the EtC can be regarded as the source coding module. During wireless transmission, the channel coding module and the modulation module should be followed to provide reliable transmission. To avoid choosing the concrete adaptive modulation and coding (AMC) strategy, the Shannon capacity\textemdash an upper bound of the transmission rate\textemdash is assumed for the EtC method to provide error-free transmission. The LE method and the PE method are originally designed for DL-based classification tasks. According to \cite{ito2021image}, the classification accuracy of ResNet-20 \cite{he2016deep} based on the LE method and the PE method are 87.02\% and 86.99\% on the CIFAR-10 test dataset, respectively. The results show that the PE method and the LE method are compatible with DL. Since the DJSCC method is also a DL-based method, we combine the PE/LE method with the DJSCC method as the compared method.

\begin{figure}[!tb]
\centering
\includegraphics[width=1\columnwidth]{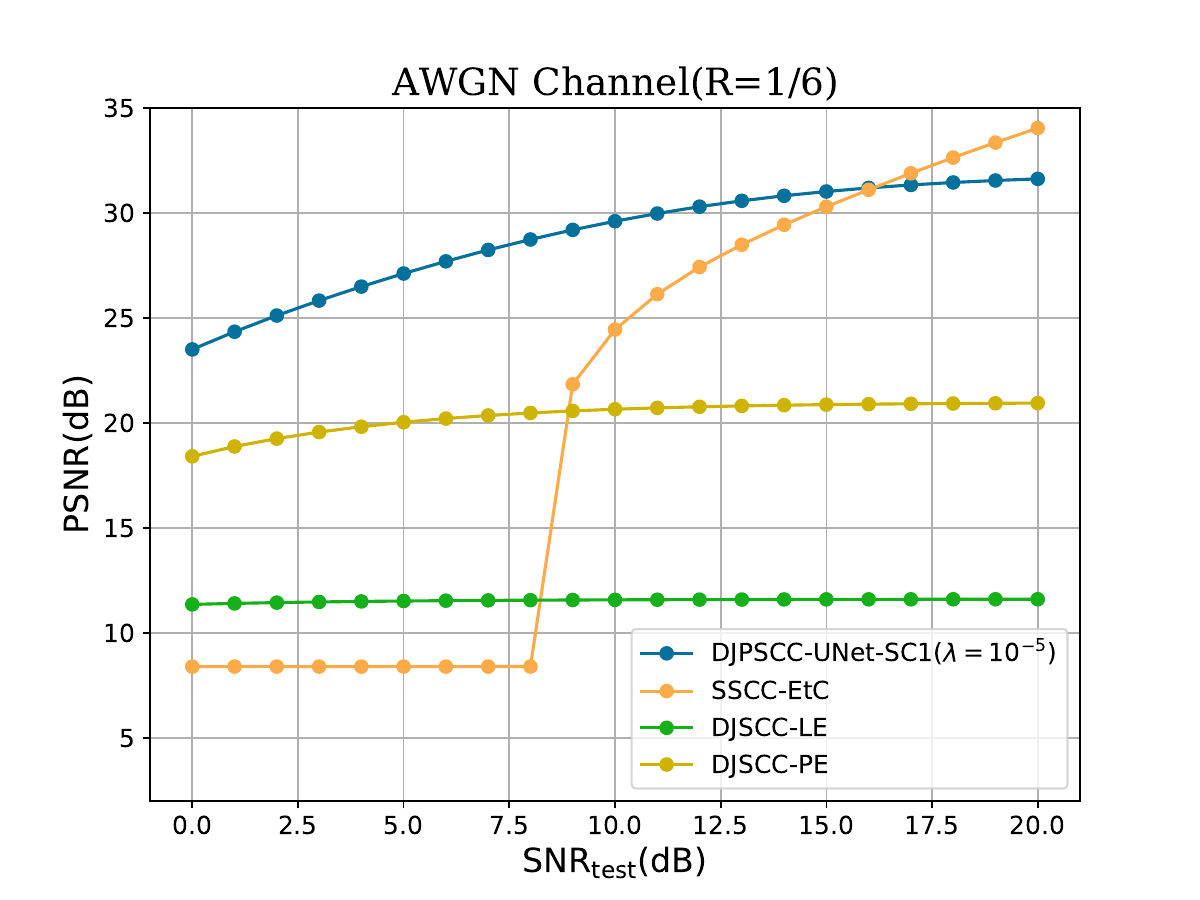}
\caption{Reconstruction performance of DJPSCC-UNet-SC1($\lambda=10^{-5}$), SSCC-EtC, DJSCC-LE and DJSCC-PE on the CIFAR-10 test dataset.}
\label{Fig:compare_cifar10_c16_snrdb0to20}
\end{figure}

Fig.~\ref{Fig:compare_cifar10_c16_snrdb0to20} compares the reconstruction performance of the DJPSCC method with that of the aforementioned methods for the bandwidth ratio $R=1/6$. The SSCC-based EtC method, the PE method combined with the DJSCC method, and the LE method combined with the DJSCC method are named as SSCC-EtC, DJSCC-PE, and DJSCC-LE, respectively. The initial flat curve of SSCC-EtC, which is around 9.5 dB when $\rm SNR_{test} \leq 8 dB$, is due to the fact that the EtC method completely breaks down in this SNR region, i.e., the minimum bit length of the EtC scheme is greater than the maximum transmission rate calculated by Shannon capacity. With the $\rm SNR_{test}$ increases from 8 dB to 16 dB, the reconstruction performance of SSCC-EtC is still worse than that of DJPSCC-UNet-SC1($\lambda=10^{-5}$), while the gap gradually decreases. When $\rm SNR_{test} \geq 16 dB$, the reconstruction performance of SSCC-EtC is superior to that of DJPSCC-UNet-SC1($\lambda=10^{-5}$). It is worth noting that DJPSCC-UNet-SC1($\lambda=10^{-5}$) is with the first DJSCC architecture proposed in \cite{bourtsoulatze2019deep}, the performance of which is inferior to that of the ADJSCC architecture proposed in \cite{xu2022wireless}. By using an advanced DJSCC architecture, the DJPSCC may have a better result and be superior to the SSCC-EtC in high SNRs. The reconstruction performance of the DJSCC-LE is around 11.5 dB when $\rm SNR_{test}$ is in the range from 0dB to 20dB, which is much lower than that of DJPSCC-UNet-SC1($\lambda=10^{-5}$). The reconstruction performance of the DJSCC-PE is much better than that of the DJSCC-LE. However, it is still 5 dB lower than the reconstruction performance of DJPSCC-UNet-SC1($\lambda=10^{-5}$) at $\rm SNR_{test}=0$ dB and the performance gap between the DJSCC-PE and DJPSCC-UNet-SC1($\lambda=10^{-5}$) is further widened with increasing SNR. 

\begin{figure}[!htb]
\centering
\includegraphics[width=0.9\columnwidth]{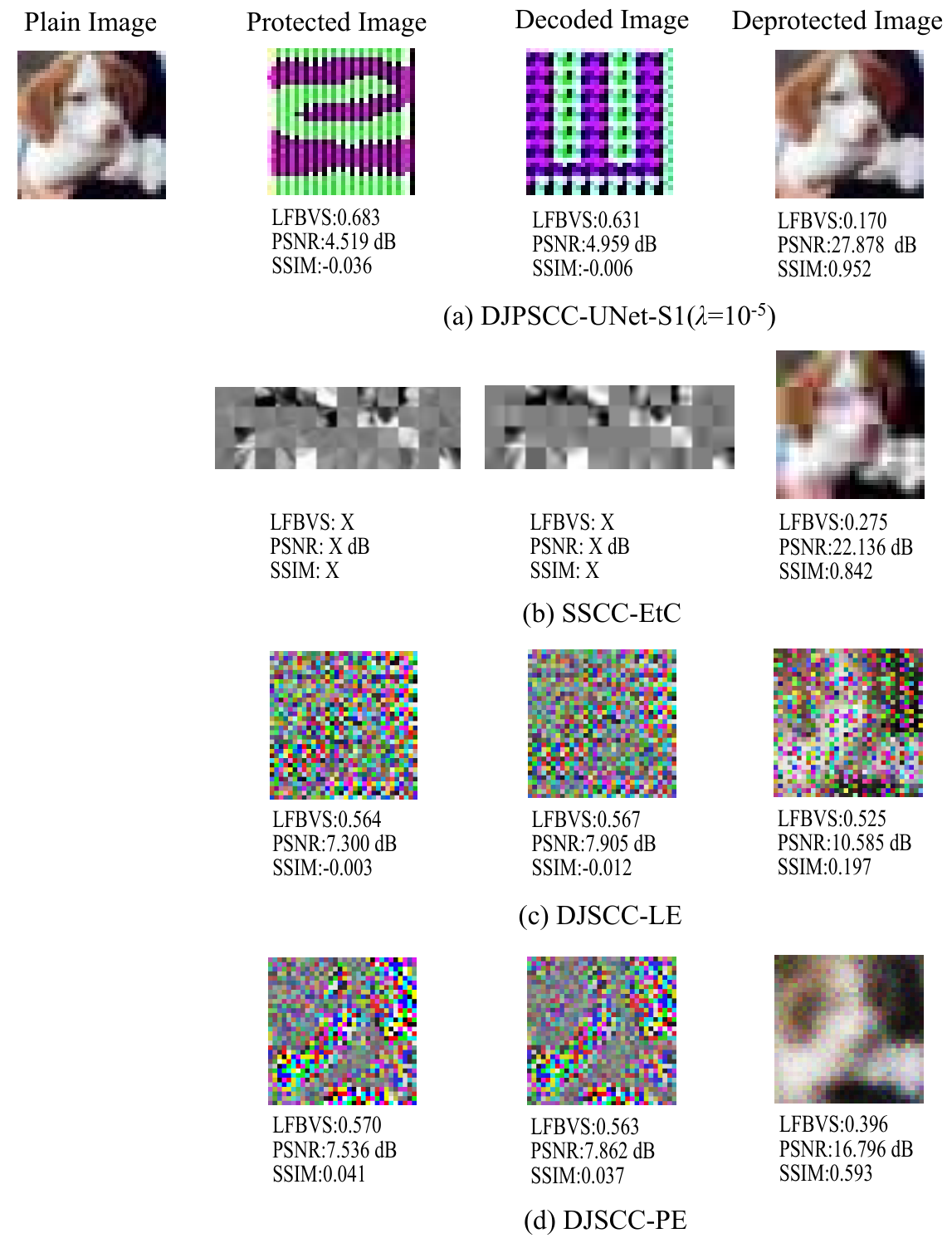}
\caption{Visually protected images comparison generated by DJPSCC-UNet-SC1($\lambda=10^{-5}$), SSCC-EtC, DJSCC-LE and DJSCC-PE with the bandwidth ratio R=1/6 and $\rm SNR$=10 dB. The image in the first column is the plain image. The images in the second column are the protected images transformed by the image owner. The images in the third column are the decoded images and the images in the last column are the deprotected images. (a) DJPSCC-UNet-SC1($\lambda=10^{-5}$), (b) SSCC-EtC, (c) DJSCC-LE, (d) DJSCC-PE.}
\label{Fig:compare_cifar10_c16_visual}
\end{figure}

Fig.~\ref{Fig:compare_cifar10_c16_visual} shows the corresponding visual performance for the aforementioned protection methods at $\rm SNR=10$ dB. Based on the EtC encryption strategy, visually protected images from SSCC-EtC are grayscale-based images with $3 \times h \times w$ pixels. Since the LFBVS, the PSNR, and the SSIM require two images with the same shape, these evaluation scores are absent for the protected image and the decoded image of the SSCC-EtC. Visually protected images of the SSCC-EtC are shown as chaotic block images, while visually protected images of the DJSCC-LE and the DJSCC-PE are shown as noisy images. All protection methods can protect the visual content of the plain image. However, when comparing the reconstruction quality of the deprotected images, DJPSCC-UNet-SC1($\lambda=10^{-5}$) shows the best performance.

\subsection{DJPSCC Generality on CIFAR-10 Dataset}
\label{generality}

We have mentioned that the DJPSCC framework is a general framework. In addition to the architecture of the UNet-SC1, other network architectures can be applied as the protection/deprotection network in the DJPSCC. Moreover, the feature extraction module can also adopt other network architectures instead of VGG16. To demonstrate the generality of the proposed method, we design new protection/deprotection architectures following the proposed principles of the protection/deprotection module and adopt a new feature extraction network for concrete DJPSCC networks to execute the following experiments.

\begin{figure}[!tb]
\centering
\includegraphics[width=1\columnwidth]{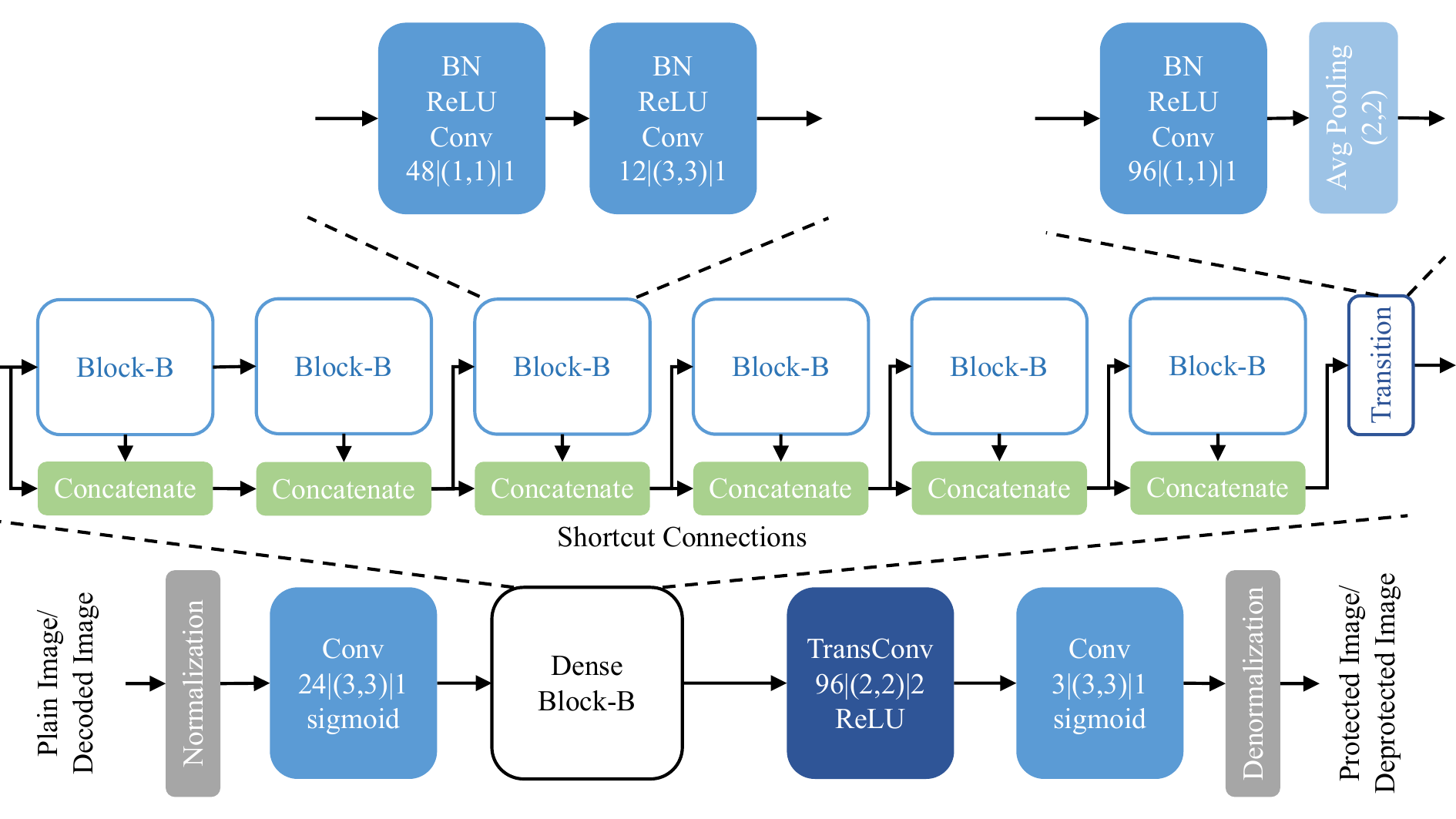}
\caption{A protection/deprotection architecture based on one dense block (DenseNet-S1).}
\label{Fig:DenseNet_C1}
\end{figure}

\begin{figure}[!tb]
\centering
\includegraphics[width=1\columnwidth]{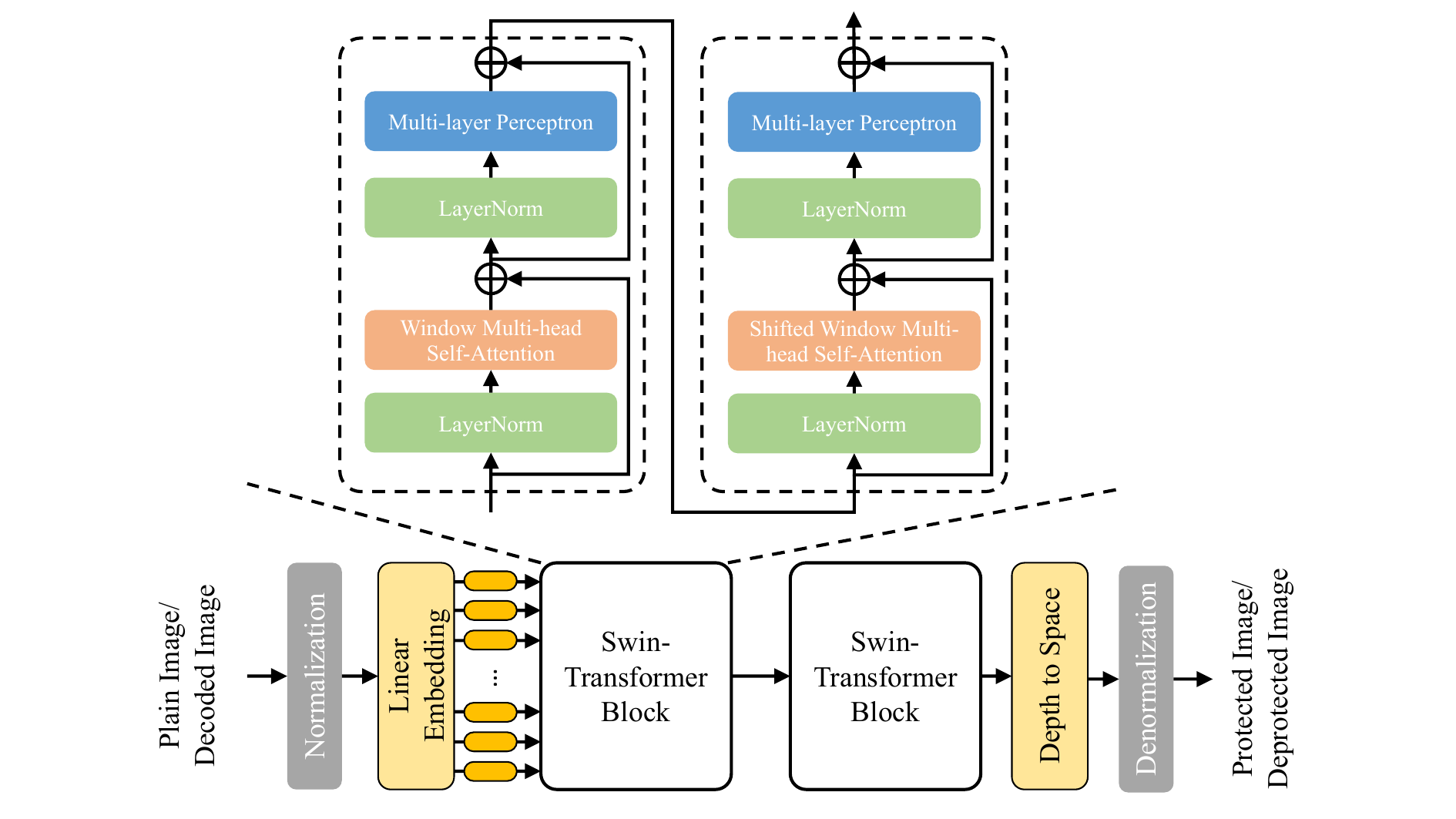}
\caption{A protection/deprotection architecture based on Swin-Transformer block (SwinTNet).}
\label{Fig:SwinTNet}
\end{figure}

The architecture of the dense block proposed in \cite{huang2017densely} uses multiple shortcut connections, satisfying the first principle about the shortcut connection. As shown in Fig.~\ref{Fig:DenseNet_C1}, we design a Dense Block-B based protection/deprotection network named as DenseNet-S1. ResNet\footnote{https://keras.io/api/applications/resnet/\#resnet50-function} \cite{he2016deep} pretrained on the ImageNet dataset is adopted as the feature extraction network. Besides the convolution neural network based architectures, different backbones, e.g., transformer-based architectures, can also be employed to verify the generality of the proposed method. To compare the influence brought by different backbones, a Swin-Transformer based protection/deprotection network illustrated in Fig.~\ref{Fig:SwinTNet} is also used as the backbone of protection/deprotection backbone. The training conditions of DJPSCC-DenseNet-S1 and DJPSCC-SwinTNet are the same as the conditions trained for DJSCC-UNet-S1.

\begin{figure}[!tb]
\centering
\includegraphics[width=1\columnwidth]{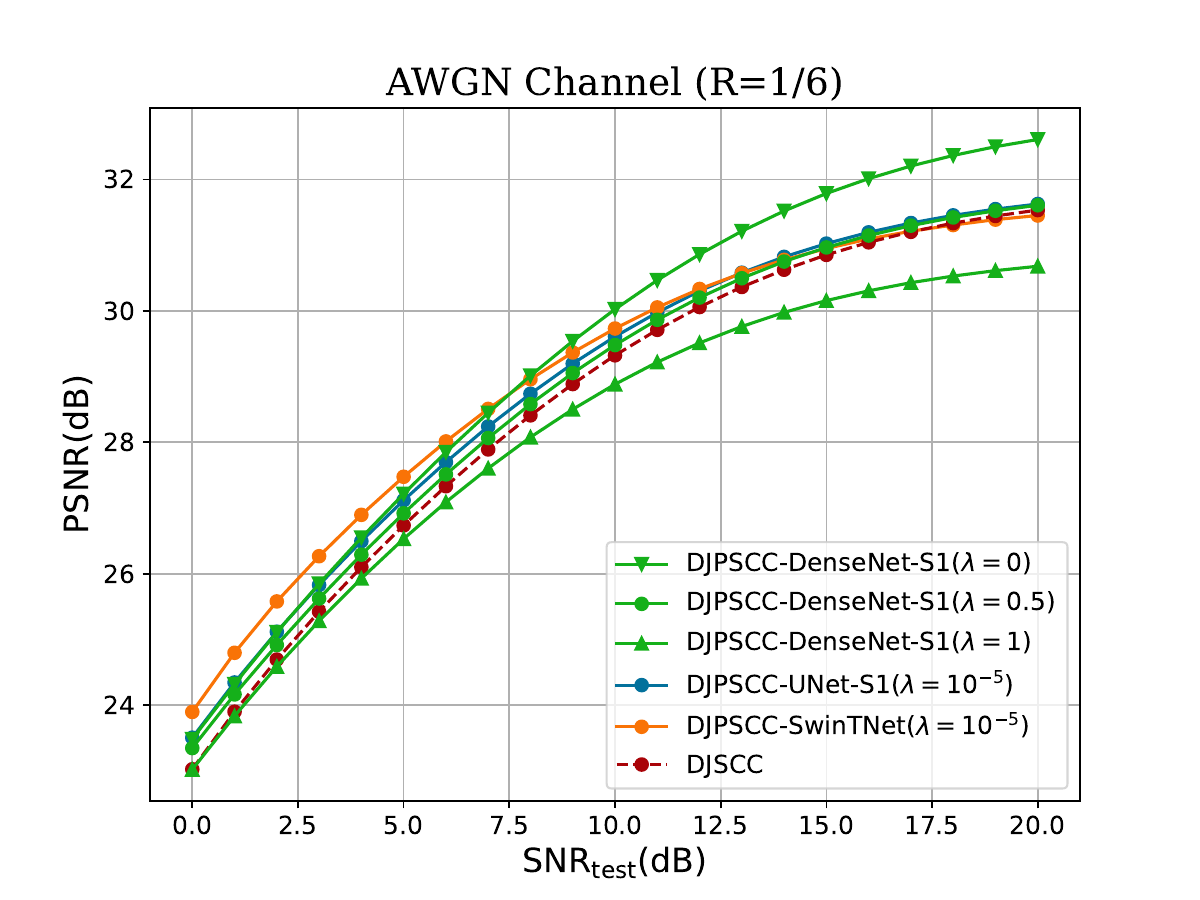}
\caption{Reconstruction performance of DJPSCC-DenseNet-S1 and DJPSCC-SwinTNet on CIFAR-10 test dataset with bandwidth ratio R=1/6.}
\label{Fig:cifar10_c16_snrdb0to20_general}
\end{figure}

\begin{table*}[!tb] 
\renewcommand{\arraystretch}{1.3} \scriptsize
\caption{Visual Security Evaluation of DJPSCC-DenseNet-S1 on the CIFAR-10 test dataset}
\label{Table:vse_densenet}
\centering
\begin{tabular}{|c|cccc|cccc|cccc|}
\hline
\multirow{2}*{\textbf{Method}} & \multicolumn{4}{c|}{\textbf{Protected Image}}& \multicolumn{4}{c|}{\textbf{Decoded Image}}&\multicolumn{4}{c|}{\textbf{Deprotected Image}}\\\cline{2-13}
~&\textbf{LFBVS}&\textbf{COR}&\textbf{PSNR(dB)}&\textbf{SSIM}&\textbf{LFBVS}&\textbf{COR}&\textbf{PSNR(dB)}&\textbf{SSIM}&\textbf{LFBVS}&\textbf{COR}&\textbf{PSNR(dB)}&\textbf{SSIM}
\\\hline
DJPSCC-DenseNet-S1 ($\lambda=0$)    & 0.573 & -0.004 & 10.514 & 0.024 & 0.592 & 0.048 & 8.829 & 0.041 & \textbf{0.202} & \textbf{0.983} & \textbf{30.023} & \textbf{0.944} \\\hline
DJPSCC-DenseNet-S1 ($\lambda=0.5$)& 0.701 & -0.004 & 6.483 & \textbf{0.006} & 0.704 & 0.000 & 5.921 & 0.004 & 0.214 & 0.981 & 29.478 & 0.937 \\\hline
DJPSCC-DenseNet-S1 ($\lambda=1$) & \textbf{0.702} & \textbf{-0.010} & \textbf{5.748} & 0.041 & \textbf{0.720} & \textbf{-0.005} & \textbf{5.598} & \textbf{0.002} & 0.224 & 0.978 & 28.879 & 0.929 \\\hline
\end{tabular}
\end{table*}
Fig.~\ref{Fig:cifar10_c16_snrdb0to20_general} and Table \ref{Table:vse_densenet} compare the reconstruction performance and the visual security performance of DJPSCC-DenseNet-S1 with the bandwidth ratio R=1/6, respectively. The reconstruction performance of DJPSCC-UNet-S1($\lambda=10^{-5}$), DJPSCC-SwinTNet($\lambda=10^{-5}$) and the DJSCC are plotted as reference in Fig.~\ref{Fig:cifar10_c16_snrdb0to20_general}. Similarly to DJPSCC-UNet-S1, with an increase of $\lambda$, the visual security performance of DJPSCC-DenseNet-S1 improves while the reconstruction performance of DJPSCC-DenseNet-S1 degrades. The reconstruction performance of DJPSCC-DenseNet-S1($\lambda=0.5$) is near that of DJPSCC-UNet-S1($\lambda=10^{-5}$). Note that $\lambda$ in DJPSCC-DenseNet-S1 is not of the same order of magnitude as $\lambda$ in DJPSCC-UNet-S1 due to the different feature extraction networks chosen in the two DJPSCC methods. Fig.~\ref{Fig:cifar10_c16_snrdb0to20_general} also provides a comparison of the reconstruction performance brought by different backbones. For the same visual security level, the reconstruction performance of DJPSCC-SwinTNet is almost the same as that of DJPSCC-UNet-S1 when $\rm SNR_{test} \geq 10 dB$. Although the reconstruction performance of DJPSCC-SwinTNet is better than that of DJPSCC-UNet-S1, the maximum reconstruction performance gap is barely less than 0.5 dB. The same results can be observed when comparing the reconstruction performance between DJPSCC-UNet-S1($\lambda=10^{-5}$) and DJPSCC-DenseNet-S1($\lambda=0.5$), if we consider that the feature extraction modules, i.e., VGG16 for DJPSCC-UNet-S1 with $\lambda=10^{-5}$ and ResNet for DJPSCC-DenseNet-S1 with $\lambda=0.5$, provide similar visual security levels. These observations may reflect the fact that the reconstruction performance of the proposed method is more dependent on the trade-off parameters $\lambda$ than the concrete protection/deprotection network architectures.

\subsection{DJPSCC Ablation Study on CIFAR-10 Dataset}
\label{ablation}

\begin{figure}[!tb]
\centering
\includegraphics[width=1\columnwidth]{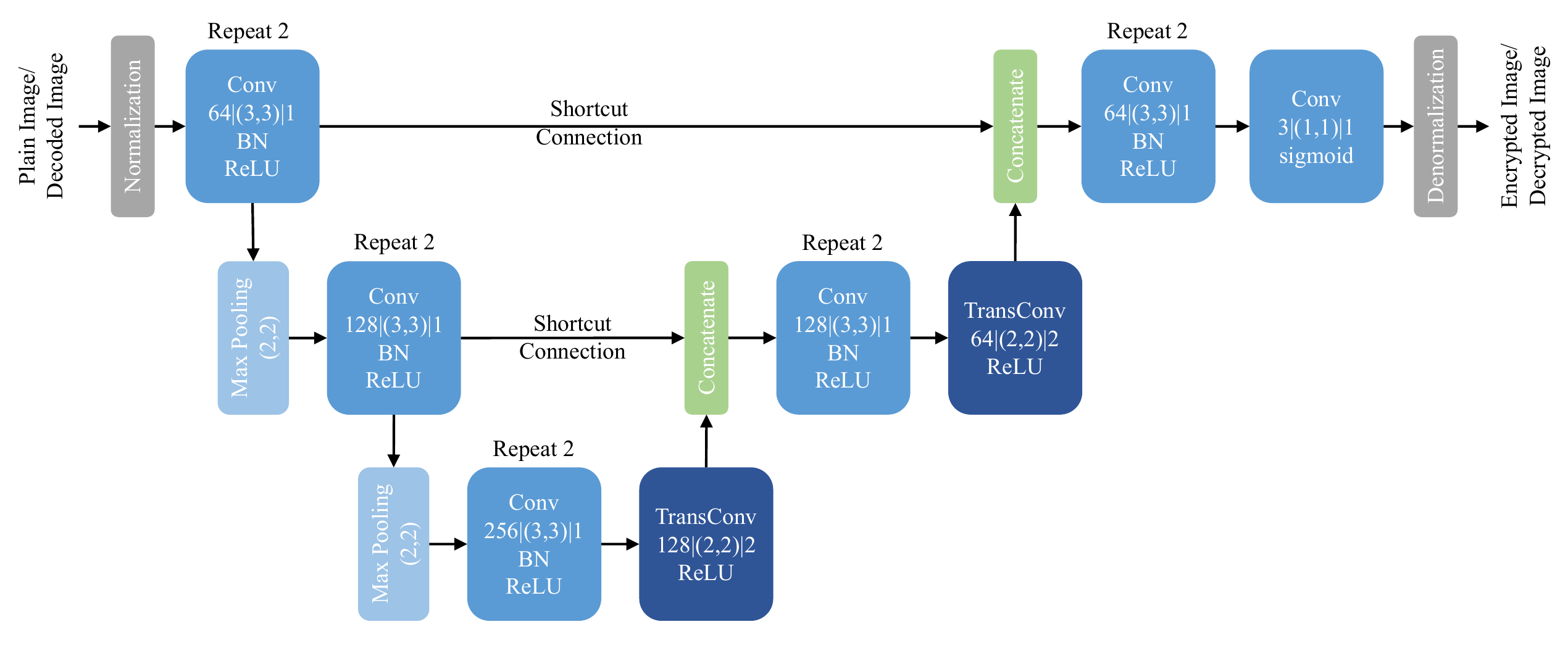}
\caption{A shallow version of U-Net with two shortcut connections (UNet-S2).}
\label{Fig:UNet_SC2}
\end{figure}

To demonstrate the efficacy of these principles in the design of the protection/deprotection module, we construct three types of protection/deprotection architectures. As shown in Fig.~\ref{Fig:UNet_SC2}, the first architecture named UNet-S2 is a deeper version of UNet-S1 with two shortcut connections. The second architecture is the UNet-S1 without the shortcut connection, which is called UNet-NS1. The UNet-S2 without shortcut connections named UNet-NS2 is the third architecture of the protection/deprotection network. Similarly, DenseNet-S2, DenseNet-NS1 and DenseNet-NS2 represent a deeper dense block based network with shortcut connections, a DenseNet-S1 without shortcut connections and a deeper network without short connections, respectively. The training conditions of the aforementioned networks are consistent with the training conditions of DJPSCC-UNet-S1 described in Section \ref{validity}. In the training stage, VGG16 and ResNet are adopted as the feature extraction network for UNet-based networks and DenseNet-based networks, respectively. To acquire similar visual protection abilities for the aforementioned methods, the loss weight $\lambda$ for UNet-based networks is set to $10^{-5}$ and that for DenseNet-based networks is set to $0.5$.

\begin{figure}[!tb]
\centering
\includegraphics[width=1\columnwidth]{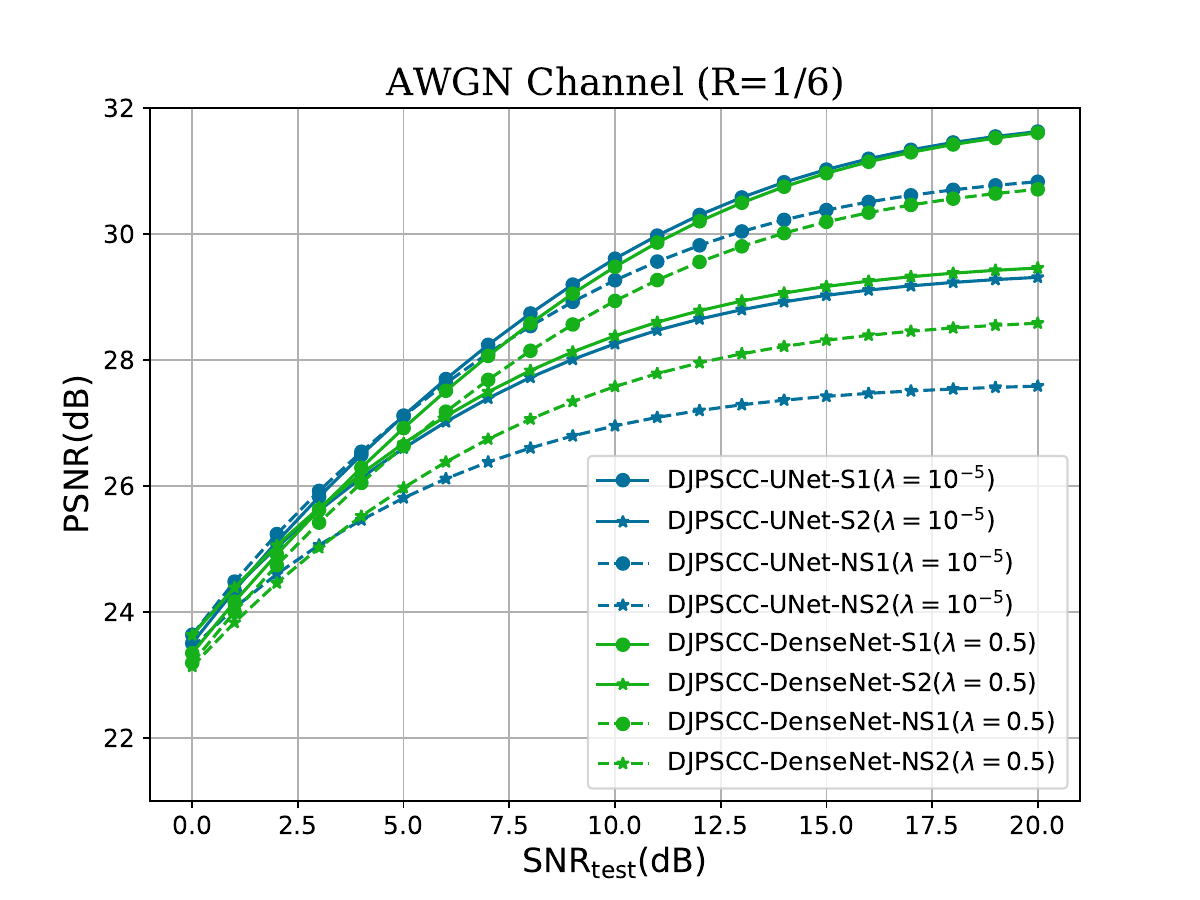}
\caption{Reconstruction performance of DJPSCC-UNet-S1, DJPSCC-UNet-S2, DJPSCC-UNet-NS1, DJPSCC-UNet-NS2, DJPSCC-DenseNet-S1, DJPSCC-DenseNet-S2, DJPSCC-DenseNet-NS1 and DJPSCC-DenseNet-NS2 evaluated on the CIFAR-10 test dataset with bandwidth ratio R=1/6.}
\label{Fig:ablation}
\end{figure}

Here, we only compare the reconstruction performance as shown in Fig.~\ref{Fig:ablation}. DJPSCC-UNet-S1 performs better than DJPSCC-UNet-NS1 at high SNRs. This performance gain comes from the shortcut connection used in DJPSCC-UNet-S1. In conventional DL-based tasks, e.g., classification tasks and semantic segmentation tasks, the shortcut connection is widely used to improve the performance. In DJPSCC, the shortcut connection in the protection/deprotection architecture has the same utility. However, the performance of DJPSCC-UNet-S2 is inferior to that of DJPSCC-UNet-S1, which is in contrast to the intuition that increasing the depth of the network can improve performance \cite{szegedy2015going}. Indeed, in conventional DL-based tasks, the shortcut connection can facilitate the training of deep networks and promote networks to learn more complex feature patterns to improve the performance. In the DJPSCC, there is a wireless channel naturally placed between the DJSCC encoder and the DJSCC decoder. It is impossible to create a shortcut connection across the wireless channel, which limits the efficiency of backward propagation in deeper networks during the training stage and causes performance degradation for deeper networks. In addition, the visual protection task and the reconstruction task executed by the proposed DJPSCC belong to low-level tasks. It is unnecessary to employ deeper protection/deprotection architectures to extract high-level semantic features. Similar results are revealed when comparing the DenseNet-based networks. Again, the shortcut connections in the protection/deprotection module can promote the reconstruction performance of the proposed DJPSCC, while deepening the protection/deprotection module brings an opposite result.

\subsection{DJPSCC Validity on Imagenet Dataset}
\label{validity_imagenet}
\begin{figure}[!tb]
\centering
\includegraphics[width=1\columnwidth]{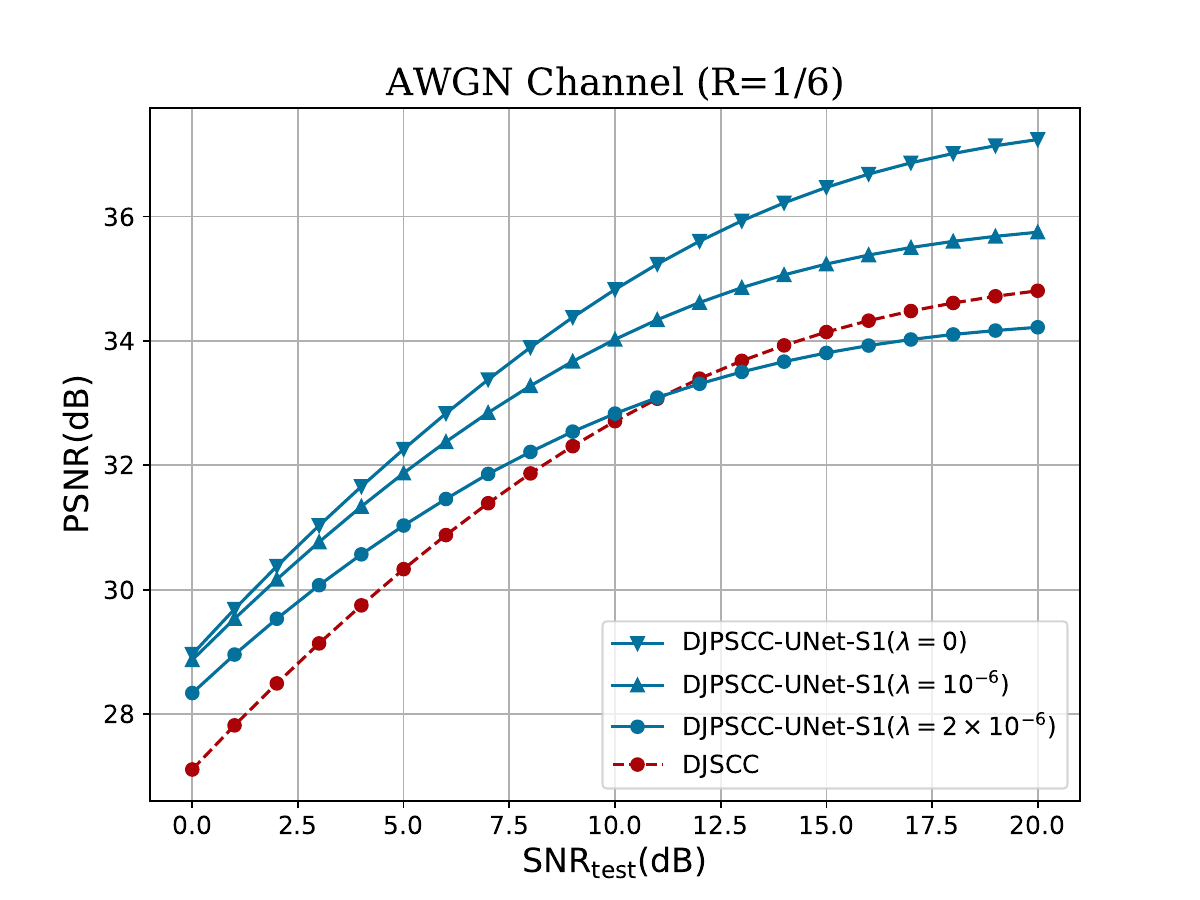}
\caption{Reconstruction performance of DJPSCC-UNet-S1 and DJSCC trained on Imagenet dataset and evaluated on Kodak dataset with R=1/6.}
\label{Fig:imagenet_c16_snrdb0to20}
\end{figure}

\begin{table*}[tb] 
\renewcommand{\arraystretch}{1.3} \scriptsize
\caption{Visual Security Evaluation of DJPSCC-UNet-S1 on the Kodak dataset}
\label{Table:vse_unet_imagenet}
\centering
\begin{tabular}{|c|cccc|cccc|cccc|}
\hline
\multirow{2}*{\textbf{Method}} & \multicolumn{4}{c|}{\textbf{Protected Image}}& \multicolumn{4}{c|}{\textbf{Decoded Image}}&\multicolumn{4}{c|}{\textbf{Deprotected Image}}\\\cline{2-13}
~&\textbf{LFBVS}&\textbf{COR}&\textbf{PSNR(dB)}&\textbf{SSIM}&\textbf{LFBVS}&\textbf{COR}&\textbf{PSNR(dB)}&\textbf{SSIM}&\textbf{LFBVS}&\textbf{COR}&\textbf{PSNR(dB)}&\textbf{SSIM}
\\\hline

DJPSCC-UNet-S1 ($\lambda=0$)    & 0.593 & \textbf{-0.041} & 12.605 & 0.167 & 0.661 & 0.025 & 6.737 & 0.008 & \textbf{0.068} & \textbf{0.994} & \textbf{34.836} & \textbf{0.946} \\\hline
DJPSCC-UNet-S1 ($\lambda=10^{-6}$)& 0.766 & -0.001 & 6.197 & 0.004 & 0.680 & \textbf{-0.023} & \textbf{6.013} & \textbf{0.002} & 0.072 & 0.993 & 34.035 & 0.939 \\\hline
DJPSCC-UNet-S1 ($\lambda=2\times 10^{-6}$) & \textbf{0.786} & -0.004 & \textbf{5.412} & \textbf{0.003} & \textbf{0.720} & -0.008 & 6.303 & 0.004 & 0.083 & 0.993 & 32.820 & 0.936 \\\hline
\end{tabular}
\end{table*}

\begin{figure*}[!htb]
\centering
\includegraphics[width=2\columnwidth]{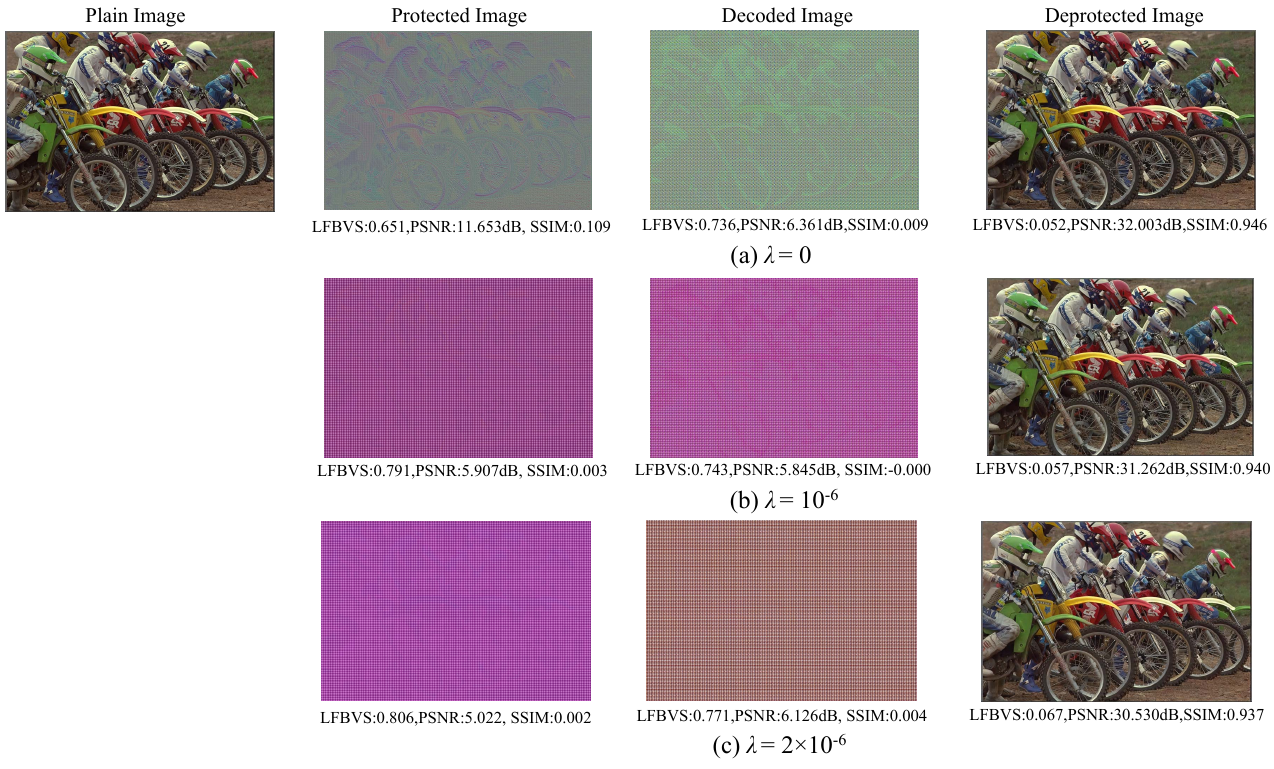}
\caption{Visually protected images generated by the proposed method at $\rm SNR=10$ dB. The image in the first column is the plain image. The images in the second column are the protected images transformed by the image owner. The images in the third column are the decoded images decoded by the DJSCC decoder. The images in the last column are the deprotected images transformed by the image receiver. (a) $\lambda=0$, (b) $\lambda=1e^{-6}$, (c) $\lambda=2\times10^{-6}$.}
\label{Fig:kodak_c16_visual}
\end{figure*}

We have demonstrated the validity of the proposed method in a low-resolution image dataset (i.e., CIFAR-10 dataset) in Section \ref{validity}. In this part, DJPSCC-UNet-S1 is trained on higher resolution image dataset (i.e.,ImageNet dataset) and evaluated on Kodak dataset. The Imagenet dataset consists of more than 1.2 million images. In the training stage, each image in the ImageNet dataset is resized to $128\times 128$ and then fed into the proposed network. Adam optimizer with a learning rate of $10^{-4}$ and a batch size of 32 are used to train the proposed model. The training process is stopped when there is no improvement in the validation loss for five consecutive epochs. In \cite{bourtsoulatze2019deep}, owing to the full convolutional architecture adopted by the DJSCC method, the Kodak dataset of size $512 \times 768$ can be directly fed into the DJSCC network and the reconstruction performance is acceptable. \cite{xu2022wireless} further demonstrates the performance of the full convolutional architecture when the test dataset is consistent/inconsistent with the training data set. Due to the full convolution network architecture of the UNet-S1 adopted in the protection/deprotection network, DJPSCC-UNet-S1 architecture is a full convolutional network and can directly deal with the Kodak dataset.

The reconstruction performance, the visual security performance and the image visualization of DJPSCC-UNet-S1 with loss weights $\lambda=0,10^{-6}, 2\times10^{-6}$ at bandwidth ratio $R=1/6$ are shown in Fig.~\ref{Fig:imagenet_c16_snrdb0to20}, Table \ref{Table:vse_unet_imagenet}  and Fig.~\ref{Fig:kodak_c16_visual}, respectively. Similarly to DJPSCC-UNet-S1 and DJPSCC-DenseNet-S1 trained on the CIFAR-10 dataset, with an increase of $\lambda$, the visual security performance of DJPSCC-UNet-S1 trained on the ImageNet dataset improves, while the reconstruction performance of DJPSCC-UNet-S1 trained on the ImageNet dataset degrades. As illustrated in Fig.~\ref{Fig:kodak_c16_visual}, although DJPSCC-UNet-S1 ($\lambda=0$) yields the best reconstruction performance, the outline of motorcycles and riders can be seen in the protected image and the decoded image. DJPSCC-UNet-S1($\lambda=10^{-6}$) protects most of the visual information. However, the shadow of motorcycles and riders can be seen vaguely in the decoded image. DJPSCC-UNet-S1($\lambda=2\times 10^{-6}$) provides the best visual protection, since we can see a regular lattice only in the protected image and the decoded image. In real wireless scenarios, if high quality of the deprotected image is required, the transmit power or bandwidth usage can be increased to improve the quality of the deprotected image of DJSCC transmissions.

\section{Conclusion}
We have proposed a novel DJPSCC method to protect the privacy and confidentiality of information. Concretely, we have constructed a general framework including the protection module, the DJSCC module, and the deprotection module, redesigned the loss function by using the feature extraction module, and proposed two principles to guide the concrete design of the protection/deprotection module. During the training stage, the proposed DJPSCC method can successfully learn an effective protection network, an effective DJSCC network, and an effective deprotection network. 

With increasing loss weight $\lambda$, the visual protection performance of the proposed DJPSCC network increases and the reconstruction performance decreases. Compared with the SSCC-based image protection method (e.g., the EtC method) and image protection methods for DL (e.g., the LE method and the PE method), the proposed DJPSCC method has shown much better reconstruction performance. The proposed DJPSCC method is a general method to protect the visual content of the plain image transmitted by DJSCC. Following the proposed design principles of the protection/deprotection module in Section \ref{Experimental Results}, multiple protection/deprotection networks can be designed to meet the requirements in real communication scenarios, e.g., storage overhead and computational complexity. In addition, with some appropriate modifications to DJPSCC, the proposed DJPSCC method can be applied to various DJSCC architectures, e.g., DJSCC-l \cite{kurka2021bandwidth}, DJSCC-f \cite{kurka2020deepjscc}, and ADJSCC \cite{xu2022wireless}.

\bibliographystyle{IEEEtran}
\bibliography{ref/ref.bib}

\begin{IEEEbiography}[{\includegraphics[width=1in,height=1.25in,clip]{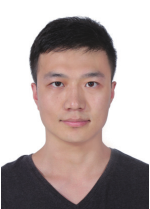}}]{Jialong Xu}
(Member, IEEE) received the B.E. and M.S. degrees from Engineering University of PAP in 2009 and 2012 respectively. He received his Ph.D. degree from Beijing Jiaotong University in 2022. He joined DOCOMO Beijing Laboratories in 2023 and is now a researcher in the solution department. His research interests include deep learning, wireless coding, information theory, and Native AI for the physical layer.
\end{IEEEbiography}

\begin{IEEEbiography}[{\includegraphics[width=1in,height=1.25in,clip]{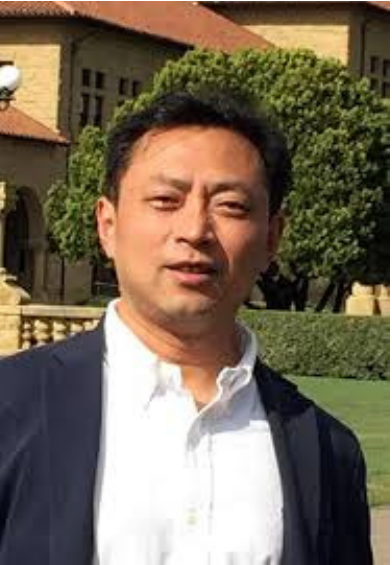}}]{Bo Ai}(Fellow, IEEE) received the M.S. and Ph.D.degrees from Xidian University, Xian, China, in 2002 and 2004, respectively. He was with Tsinghua University, Beijing, China, where he was an Excellent Postdoctoral Research Fellow in 2007. He is currently a Professor and an Advisor of Ph.D.candidates with Beijing Jiaotong University, Beijing, where he is also the Deputy Director of the State Key Laboratory of Rail Traffic Control and Safety. He is also currently with the Engineering College, Armed Police Force, Xian. He has authored or coauthored six books and 270 scientific research papers, and holds 26 invention patents in his research areas. His interests include the research and applications of orthogonal frequency-division multiplexing techniques, high-power amplifier linearization techniques, radio propagation and channel modeling, global systems for mobile communications for railway systems, and long-term evolution for railway systems.

Dr. Ai is a Fellow of The Institution of Engineering and Technology. He was as a Co-chair or a Session/Track Chair for many international conferences such as the 9th International Heavy Haul Conference (2009); the 2011 IEEE International Conference on Intelligent Rail Transportation; HSRCom2011; the 2012 IEEE International Symposium on Consumer Electronics; the 2013 International Conference on Wireless, Mobile and Multimedia; IEEE Green HetNet 2013; and the IEEE 78th Vehicular Technology Conference (2014). He is an Associate Editor of IEEE TRANSACTIONS ON CONSUMER ELECTRONICS and an Editorial Committee Member of the Wireless Personal Communications journal. He has received many awards such as the Qiushi Outstanding Youth Award by HongKong Qiushi Foundation, the New Century Talents by the Chinese Ministry of Education, the Zhan Tianyou Railway Science and Technology Award by the Chinese Ministry of Railways, and the Science and Technology New Star by the Beijing Municipal Science and Technology Commission.
\end{IEEEbiography}

\begin{IEEEbiography}[{\includegraphics[width=1in,height=1.25in,clip]{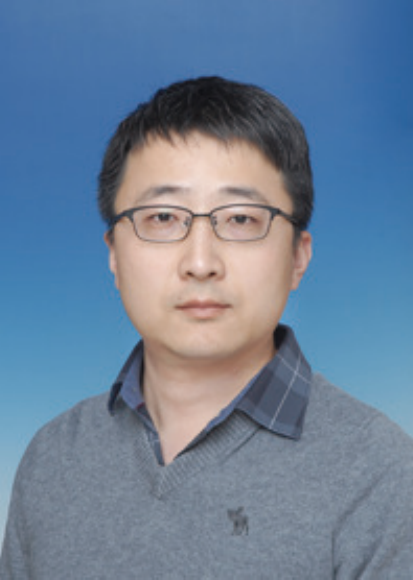}}]{Wei Chen}
(Senior Member, IEEE) received the B.Eng. and M.Eng. degrees in communications engineering from the Beijing University of Posts and Telecommunications, Beijing, China, in 2006 and 2009, respectively, and the Ph.D. degree in computer science from the University of Cambridge, Cambridge, U.K., in 2013. He was a Research Associate with the Computer Laboratory, University of Cambridge from 2013 to 2016. He is currently a Professor with Beijing Jiaotong University, Beijing. His current research interests include sparse representation, Bayesian inference, wireless communication systems and image processing. He was the recipient of the 2013 IET Wireless Sensor Systems Premium Award and the 2017 International Conference on Computer Vision (ICCV) Young Researcher Award.
\end{IEEEbiography}

\begin{IEEEbiography}[{\includegraphics[width=1in,height=1.25in,clip]{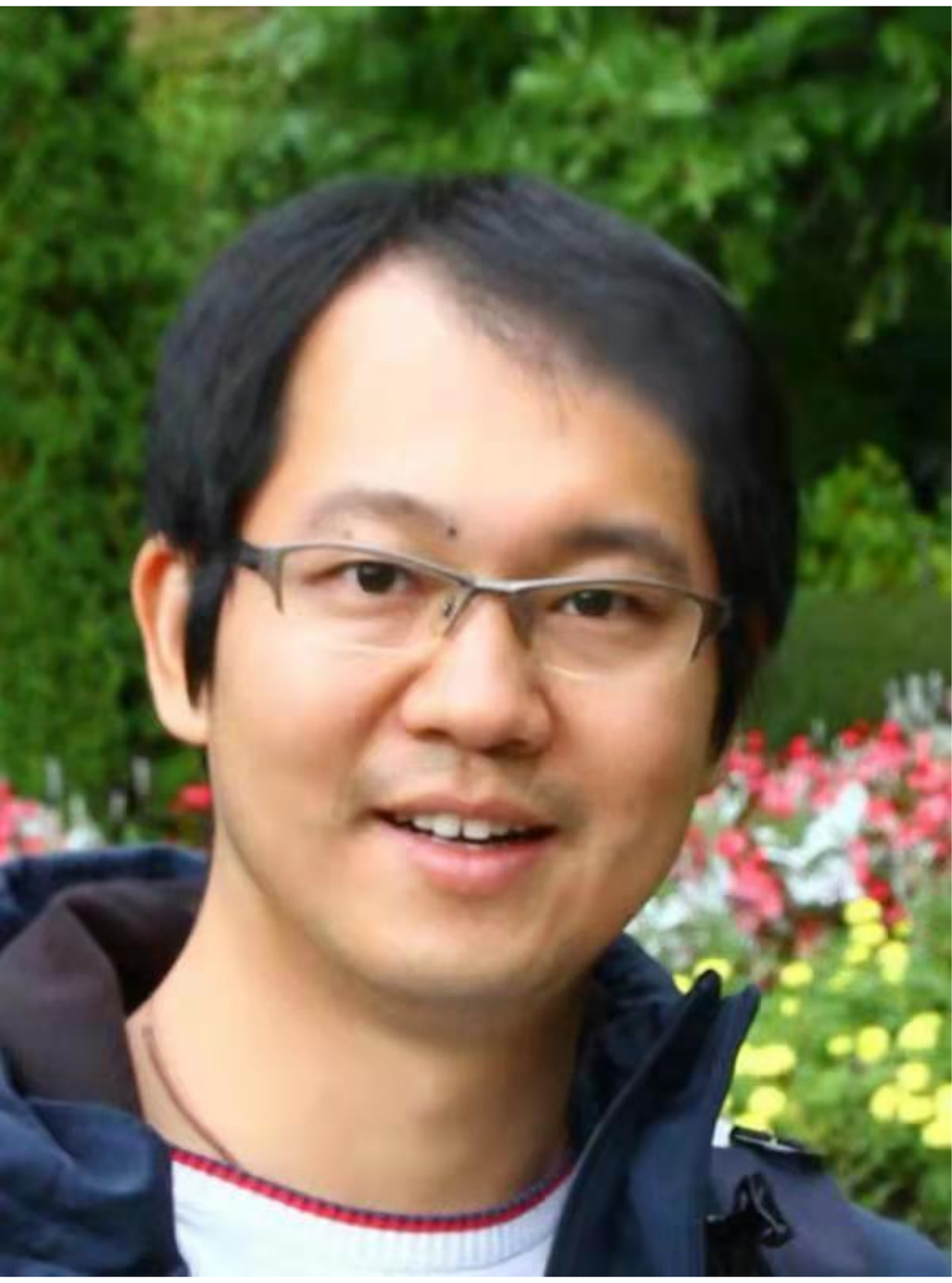}}]{Ning Wang}
(Member, IEEE) received the B.E. degree in communication engineering from Tianjin University, China, in 2004, the M.A.Sc. degree in electrical engineering from The University of British Columbia, Canada, in 2010, and the Ph.D. degree in electrical engineering from the University of Victoria, Canada, in 2013. He was on the Finalist of the Governor General's Gold Medal for Outstanding Graduating Doctoral Student with the University of Victoria in 2013. 
From 2004 to 2008, he was with the China Information Technology Design and Consulting Institute as a Mobile Communication System Engineer, specializing in planning and design of commercial mobile communication networks, network traffic analysis, and radio network optimization. He was a Postdoctoral Research Fellow of the Department of Electrical and Computer Engineering with The University of British Columbia, from 2013 to 2015. Since 2015, he has been with the School of Information Engineering, Zhengzhou University, Zhengzhou, China, where he is currently an Associate Professor. He also holds adjunct appointments with the Department of Electrical and Computer Engineering, McMaster University, Hamilton, Canada, and the Department of Electrical and Computer Engineering, University of Victoria, Victoria, Canada. He has served on the technical program committees of international conferences, including the IEEE GLOBECOM, IEEE ICC, IEEE WCNC, and CyberC. His research interests include resource allocation and security designs of future cellular networks, channel modeling for wireless communications, statistical signal processing, and cooperative wireless communications.
\end{IEEEbiography}

\begin{IEEEbiography}[{\includegraphics[width=1in,height=1.25in,clip]{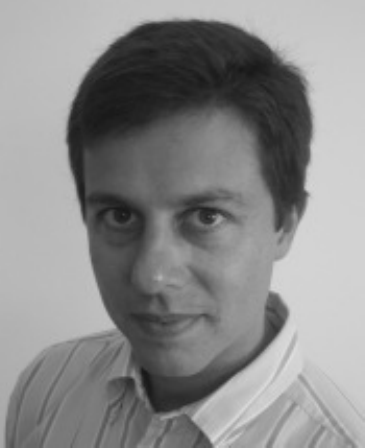}}]{Miguel Rodrigues}
(Fellow, IEEE) received the Licenciatura degree in electrical and computer engineering from the University of Porto, Porto, Portugal, and the Ph.D. degree in electronic and electrical engineering from the University College London (UCL), London, U.K. He is currently a Professor of Information Theory and Processing, UCL, and a Turing Fellow with the Alan Turing Institute - the UK National Institute of Data Science and Artificial Intelligence. His research lies in the general areas of information theory, information processing, and machine learning. His work has led to more than 200 articles in leading journals and conferences in the field, a book on Information-Theoretic Methods in Data Science (Cambridge Univ. Press), and the IEEE Communications and Information Theory Societies Joint Paper Award 2011. He is an Associate Editor for the IEEE TRANSACTIONS ON INFORMATION THEORY, and the IEEE OPEN JOURNAL OF THE COMMUNICATIONS SOCIETY. He was an Associate Editor for the IEEE COMMUNICATIONS LETTERS, and a Lead Guest Editor of the Special Issue on “Information-Theoretic Methods in Data Acquisition, Analysis, and Processing” of the IEEE JOURNAL ON SELECTED TOPICS IN SIGNAL PROCESSING. He was a Co-Chair of the Technical Programme Committee of the IEEE Information Theory Workshop 2016, Cambridge, U.K. He is a member of the IEEE Signal Processing Society Technical Committee on “Signal Processing Theory and Methods”, and the EURASIP SAT on Signal and Data Analytics for Machine Learning (SiG-DML).
\end{IEEEbiography}

\end{document}